\documentclass[final,12pt,authoryear]{elsarticle}

\usepackage{amssymb}
\usepackage{amsmath}
\usepackage{subfigure}
\usepackage[T1]{fontenc}
\usepackage{xcolor}
\usepackage{hyperref}
\usepackage{soul}
\usepackage{tikz}
\usetikzlibrary{shapes,arrows}

\journal{journal}

\begin{document}

\begin{frontmatter}



\title{upstreamFoam: an OpenFOAM-based solver for heterogeneous porous media at different scales}


\author[inst1]{Roberto Lange}
\ead{roberto.lange@wikki.com.br}

\author[inst1]{Gabriel M. Magalhães}
\ead{gabriel.magalhaes@wikki.com.br}

\author[inst1]{Franciane F. Rocha\corref{mycorrespondingauthor}}
\ead{franciane.rocha@wikki.com.br}
\cortext[mycorrespondingauthor]{Corresponding author}

\author[inst1]{Pedro V. S. Coimbra}
\ead{pedro.coimbra@wikki.com.br}

\author[inst1]{Jovani L. Favero}
\ead{jovani.favero@wikki.com.br}

\author[inst1]{Rodrigo A. C. Dias}
\ead{rodrigo.dias@wikki.com.br}

\author[inst1]{Antonio O. S. Moraes}
\ead{antonio.samel@wikki.com.br}

\author[inst2]{Mateus P. Schwalbert}
\ead{mateusps@petrobras.com.br}

\affiliation[inst1]{organization={WIKKI Brasil LTDA},
            addressline={Rua Alo\'isio Teixeira, 278, Pr\'edio 3 - sala 301, Ilha da Cidade Universit\'aria}, 
            city={Rio de Janeiro},
            postcode={21941-850}, 
            state={RJ},
            country={Brazil}}

\affiliation[inst2]{organization={Petrobras Research Center (CENPES)},
            addressline={Av. Horácio Macedo 950, Cidade Universitária, Ilha do Fundão}, 
            city={Rio de Janeiro},
            postcode={21941-598}, 
            state={RJ},
            country={Brazil}}

\begin{abstract}
\textcolor{blue}{This is the preprint version of the published manuscript \url{https://doi.org/10.1016/j.ijmultiphaseflow.2024.104954}. Please cite as:\vspace*{0.2cm}
\newline
Lange, R.; Magalhães, G.M.; Rocha, F.F.; Coimbra, P.V.S.; Favero, J.L.; Dias, R.A.; Moraes, A.O.S; Schwalbert, M.P., 2024. Development of a new computational solver for multiphase flows in heterogeneous porous media at different scales. International Journal of Multiphase Flow, 104954.
}
\bigskip
\newline
This work presents the development of a novel solver tailored for simulating multiphase flows within heterogeneous porous media. Leveraging the Eulerian multi-fluid model coupled with Darcy's law, the solver demonstrates adaptability across diverse scales, effectively handling heterogeneous porosity and permeability fields. The proposed solver, called \textit{upstreamFoam}, extends the capabilities of OpenFOAM framework, specifically the \textit{multiphaseEulerFoam}, by incorporating models for porous media simulations. This integration introduces new features and formulations, allowing for the simulation of compressible multiphase flows in porous media with intricate properties. The approach presented here provides a robust framework for characterizing reservoirs and treating heterogeneous porous systems at different scales. A successful validation of the introduced solver for classical problems with analytical, semi-analytical, and reference solutions is presented. Then, applications on a wide range of multiphase flows in heterogeneous porous media at different scales have been studied, demonstrating the potential of the solver to simulate complex multiphase problems.
\end{abstract}

%

\begin{keyword}
Multiphase ﬂow \sep Porous media \sep Finite volume \sep OpenFOAM
\end{keyword}

\end{frontmatter}




\section{Introduction}
\label{sec:introduction}

Multiphase flows describe several problems with applications in various fields. The Computational Fluid Dynamics (CFD) approaches have been extensively used for multiphase ﬂows modeling and simulation \citep{spalding1981numerical}. Implementation of methodologies for solving CFD problems within open-source simulators is gaining popularity in industrial and academic research groups. The OpenFOAM library is an example, which is based on the Finite Volume Method (FVM) and covers a variety of cases \citep{jasak1996error, weller1998tensorial}. 

A well-known procedure for solving CFD problems is the Eulerian multi-fluid model \citep{hill1998computer, ishii2010thermo}, which describes a system of phase fractions represented by averaged conservation equations \citep{crowe1998multiphase, drew2006theory}. In this context, work of \cite{weller2002derivation} should be highlighted, where a conservative and bounded formulation based on a volumetric average for the momentum exchange terms has been presented.

There are different solution procedures for multiphase flow problems available in OpenFOAM. In the framework of Eulerian fluid models, the solver \textit{twoPhaseEulerFoam} has been proposed, initially to simulate incompressible two-phase flows \citep{rusche2003computational}. \cite{silva2008implementation} included the effects of particle-particle interaction by coupling the population balance equation with the \textit{twoPhaseEulerFoam}. Then, the solver was extended to $n+1$ phases, considering a continuous phase and $n$ dispersed phases \citep{silva2011development, favero2015modeling}. Another solver, the \textit{multiphaseEulerFoam}, has been developed to combine the Eulerian multi-fluid solution framework with the Volume of Fluid (VOF) method for interface capturing on selected phase pairs \citep{wardle2013hybrid}. The proposed hybrid multiphase CFD solver employs the interface compression methodology of \cite{weller2008new}. The \textit{multiphaseEulerFoam} considers a system of any number of compressible fluid phases with a common pressure, but with other separate properties. This approach allows for solving complex flows with multiple species, heat and mass transfer, being widely used \citep{tocci2016assessment}.

A wide range of multiphase flow applications occurs within porous media \citep{bear1988dynamics}. The subsurface flows in oil reservoirs are relevant examples of such topics \citep{aziz1979petroleum}. In the last years, several computational models for simulating multiphase flows in porous media have been introduced \citep{chen2006computational}. An increase in the development of open-source simulators for this class of problems is also noticeable.

Regarding the OpenFOAM approach, a remarkable toolbox to simulate incompressible two-phase flows in porous media, called \textit{porousMultiphaseFoam}, has been introduced by \cite{horgue2015open}. This development includes a formulation for the phase saturations, relative permeability and capillary models, and special boundary conditions. Extensions of the \textit{porousMultiphaseFoam} include an approach for adaptive mesh refinement \citep{sugumar2020grid} and a generalization for the black oil model \citep{fioroni2021openfoam}. Recent developments have been presented by \cite{horgue2022porousmultiphasefoam}, such as improved numerical techniques for hydrogeological ﬂows modeling and the transport problem of a mixture with $n$ components. The \textit{porousMultiphaseFoam} has also been used as the basis for the development of a hybrid approach to simulate two-phase ﬂows in systems containing solid-free regions and porous matrices \citep{carrillo2020multiphase}, besides that a two-phase flow procedure coupled with geomechanics and Embedding Discrete Fracture Model (EDFM) \citep{sangnimnuan2021development}.

In this paper, we develop a solver for multiphase flows in porous media combining the Eulerian multi-fluid formulation for a system of phase fractions with the Darcy’s law for flows through porous media \citep{muskat1981physical}. The mathematical model considers a locally averaged Navier-Stokes equation \citep{weller2002derivation, rusche2003computational} with an additional momentum source term, the Darcy term, considered to represent the porous domain and the effects related to permeability field, capillarity, and viscous forces.  Since this type of modified Navier-Stokes system avoids the explicit treatment of the boundary conditions at the fluid and porous interface, it has been extensively used to simulate porous media flows \citep{goyeau2003momentum}. We can mention other related approaches, generally used to solve ﬂow problems through porous and solid-free regions simultaneously, such as the Darcy-Brinkman model \citep{brinkman1949calculation}, that uses a spatially dependent penalization term within the Navier-Stokes fluid momentum equation. More recently, a hybrid scale Darcy-Brinkman-based modeling framework, called micro-continuum approach, has been presented by \cite{soulaine2016micro}, and then extended by \cite{carrillo2020multiphase}. Another usual strategy is the definition of a mask function to set both porous areas with Darcy and Forchheimer coefficients \citep{ph1901wasserbewegung}, and free areas where the classical Navier–Stokes momentum equation is solved. It is also very common to find works in the literature that solve mass conservation equations for each fluid with velocities expressed by Darcy's law instead of solving a modified Navier-Stokes equation \citep{wu2015multiphase}. However, we choose to use a more complete set of governing equations that may benefit future extensions.

The implementation of the proposed approach, in OpenFOAM, is based on the very used \textit{multiphaseEulerFoam}, that has the capability to simulate a system of any number of compressible phase fractions at a common pressure, with the possibility of adding multiple species and mass transfer. We included the Darcy term for porous media modeling to the classic Navier-Stokes equation considered in the \textit{multiphaseEulerFoam}, enabling it to be used for applications in porous media systems. In this new scenario, specialized models for reservoir simulation, such as those considered in \textit{porousMultiphaseFoam}, are needed. We take advantage of the treatment for the relative permeability, capillarity models, and specific boundary conditions introduced by the \textit{porousMultiphaseFoam}. Additionally, we incorporate new features and formulations, such as heterogeneous porosity fields, compressibility models, the definition of stationary phases, and the possibility of using more than two moving phases. Hence, the proposed solver, called \textit{upstreamFoam}, extends the capabilities of the \textit{multiphaseEulerFoam} and \textit{porousMultiphaseFoam} to a greater variety of applications. A distinguishing property of the method presented here is the formulation of the solid as a stationary phase, which allows for the use of different minerals or different types of solids. With the new toolbox, we can handle complex simulations of compressible multiphase flows in porous media with intricate properties, which cannot be simulated independently in the previous solvers. Comparisons between \textit{upstreamFoam} and \textit{porousMultiphaseFoam} are presented in the numerical section for some porous media problems.

The multiphase porous media model requires the solution of a system of coupled nonlinear partial differential equations. The numerical formulation adopted approximates the solution by a sequential scheme based on IMplicit Pressure Explicit Saturation (IMPES) \citep{coats2000note}, where the global mass conservation, expressed by a pressure equation, is implicitly solved, while the phase fractions are explicitly solved. More specifically, the segregated algorithm considered applies an external Pressure IMplicit with splitting of operator for Pressure-Linked Equations (PIMPLE) loop for pressure-velocity coupling \citep{wang2018literature} at each time step to approximate the problems of flow and transport.  Inside the PIMPLE loop, the phase fractions are solved explicitly using the Multidimensional Universal Limiter with Explicit Solution (MULES) \citep{damian2014extended}, an OpenFOAM implementation of the Flux Corrected Transport (FCT) theory \citep{rudman1997volume}. Due to the explicit treatment, the time steps need to satisfy the Courant-Friedrichs-Lewy (CFL) condition \citep{courant1928partiellen}. In the numerical formulation section, we show two classic criteria for the time step choice implemented in \textit{upstreamFoam}: the Courant condition \citep{courant1928partiellen}, and the Coats restriction \citep{coats2003impes}. Furthermore, details on reservoir characterization are highlighted, the relationship between the formulation based on phase fractions and other common approaches based on saturations is discussed, and the models considered for relative permeability, capillary pressure, and compressibility are presented.

The paper is divided into sections as follows. 
In section \ref{sec:mathematical_model}, the mathematical model is presented, followed by the numerical formulation in section \ref{sec:numerical_formulation}, numerical results in section \ref{sec:numerical_results}, and conclusions in section \ref{sec:conclusion}.

\section{Mathematical Model}
\label{sec:mathematical_model}

The Eulerian multi-fluid model considered in this work is given by volume averaged Navier-Stokes equations as presented by \cite{weller2002derivation} and \cite{rusche2003computational}. Since the main purpose of this work is to develop an application in OpenFOAM for multiphase flows in porous media to be used on oil and gas industry simulations, an additional Darcy term is used to account the porous media effects. Therefore, the mathematical formulation is presented with a particular emphasis on incorporating treatment for porous media.

The mathematical model considers that each volume contains both solid and void space. The void space, in turn, can contain multiple fluids. In the next subsections, porous media definitions and conservation principles that describe flows in porous media are presented.

\subsection{Porous media definitions}
\label{sub:porous_definitions}

The void space of a porous medium is represented by the porosity field
\begin{equation}
\alpha_v=\frac{V_{v}}{V},
\label{eq:porosity}
\end{equation}
where $V_{v}$ is the volume occupied by the void space and $V$ is the volume of the respective cell, that is, the porosity is the void fraction and satisfies $0\leq\alpha_v\leq 1$, where $\alpha_v=0$ indicates an impermeable medium and $\alpha_v=1$ a totally free region. We consider a system of phase fractions $\alpha_i$, in which the index $i$ represents one phase of the flow, such that for each cell
\begin{equation}
\alpha_i=\left\{
\begin{array}{ll}
0,  &  \mbox{cell does not contain phase } i,\\
]0,1[\, , & \mbox{cell contains phase } i \mbox{ and other phases},\\
1,  & \mbox{cell is completely filled with phase } i.
\end{array}
\right.
\label{eq:phase_possibilities}
\end{equation}
Phase fractions satisfy the following relationship
\begin{equation}
\sum_{i=1}^{N_s} \alpha_i+\sum_{i=1}^{N_m} \alpha_i=1,
\label{eq:closure}
\end{equation}
where $N_s$ is the number of stationary phases and $N_m$ is the number of moving phases of the system. The total number of phases is denoted by $N=N_s+N_m$. In terms of the void fraction we have
\begin{equation}
\alpha_v=1-\sum_{i=1}^{N_s}\alpha_i.
\label{eq:void}
\end{equation}
It is important to notice that the solid is treated as a stationary phase, allowing to define different minerals or different types of solid as different stationary phases. 
Another advantage of this formulation is the possibility of considering residual saturations or particulate obstructions as stationary phases inclusively, allowing for the mass transfer between phases. In a numerical view, the solution of stationary phases is cheaper than the solution of moving phases.

Formulations based on saturations are more common than the phase fractions approach, however, the relationship between both consists in the fact that the saturation of a phase can be written as 
\begin{equation}
S_i=\frac{\alpha_i}{\alpha_v}.
\label{eq:saturation}
\end{equation}

\subsection{Balance equations}
\label{sub:darcy}

In the context of porous media flows, the phase momentum balance equation of each phase fraction is given by:
\begin{equation}
 \dfrac{\partial(\alpha_i\rho_i\mathbf{u}_i)}{\partial t} + \nabla\cdot(\alpha_i \rho_i \mathbf{u}_i\mathbf{u}_i)= -\alpha_i\nabla p_i + \alpha_i\rho_i\mathbf{g}  - \alpha_i^2 \frac{\mu_i}{k_{r,i}}\, \mathbf{K}^{-1} \mathbf{u}_i + \mathbf{F}_i,
	\label{eq:momentum}
\end{equation}
in which, $\mathbf{u}_i$ is the velocity of the phase $i$, $\rho_i$ is the density of phase, $p_i$ is the pressure of the phase, $\mathbf{g}$ is the gravity acceleration, $\mu_i$ is the dynamic viscosity of the phase, $k_{r,i}$ is the phase relative permeability, $\mathbf{K}$ is the absolute permeability tensor, and $\mathbf{F}_i$ is a source term to account for additional effects. The above equation represents a complete momentum balance equation, that can consider turbulent stress, interfacial forces, and surface tension through the term $\mathbf{F}_i$ \citep{weller2002derivation, rusche2003computational}. Worth mentioning that the Darcy term containing permeability effects is dominant in the cases studied here \citep{muskat1981physical}. Moreover, we highlight that the current formulation share similarities with the general modeling framework based on Darcy-Brinkman-Stokes equation presented by \cite{soulaine2016micro}.

The pressure gradient $\nabla p_i$ can be replaced using a modified pressure $p_i^*$ as follows:
\begin{equation}
  \nabla p_i = \nabla p_i^*  + \nabla \left(\rho_m (\mathbf{g}\cdot \mathbf{h})\right),
\label{eq:deltaP}
\end{equation}
where the term $\rho_m (\mathbf{g}\cdot \mathbf{h})$ represents a pseudo-hydrostatic pressure, $\mathbf{h}$ is the height of the fluid particle, and $\rho_m$
is the fluid mixture density, defined as
\begin{equation}
  \rho_m = \sum_{i=1}^{N_m} \alpha_i\rho_i.
\label{eq:rho_m}
\end{equation}
Substituting Eq. \eqref{eq:deltaP} in Eq. \eqref{eq:momentum} and
rearranging, yields:
\begin{equation}
\begin{aligned}
 \dfrac{\partial(\alpha_i\rho_i\mathbf{u}_i)}{\partial t} + \nabla\cdot(\alpha_i \rho_i \mathbf{u}_i\mathbf{u}_i)= -\alpha_i\nabla p^*_i &+ \alpha_i(\rho_i - \rho_m)\mathbf{g} - \alpha_i(\mathbf{g}\cdot \mathbf{h}) \nabla \rho_m  \\
 &- \alpha_i^2 \frac{\mu_i}{k_{r,i}}\, \mathbf{K}^{-1} \mathbf{u}_i + \mathbf{F}_i.
 \end{aligned}
\label{eq:modified_momentum}
\end{equation}

The mass balance equation for each phase reads:
\begin{equation}
  \frac{\partial(\alpha_i \rho_i)}{\partial t} + \nabla \cdot
  (\alpha_i\rho_i\mathbf{u}_i) = q_i,
\label{eq:continuity}
\end{equation}
where $q_i$ is the source term related to the phase $i$.  Equation \eqref{eq:continuity} can be expanded and rewritten as
\begin{equation}
\frac{\partial \alpha_i}{\partial t}+\nabla \cdot\left(\alpha_i \mathbf{u}_i\right)=\frac{q_i}{\rho_i}-\frac{\alpha_i}{\rho_i} \frac{D_i \rho_i}{D t},
\end{equation}
where the material derivative of $\rho$, which accounts for the compressibility effects, is given by
\begin{equation}
  \frac{D\rho_i}{Dt} = \frac{\partial \rho_i}{\partial t} + \mathbf{u}_i \cdot
  \nabla \rho_i.
\label{eq:material_derivative}
\end{equation}

Another important concept is the mixture velocity $\mathbf{u}_m$  which, for $N$ phases, is given by
\begin{equation}
  \mathbf{u}_m = \sum_{i=1}^{N} \alpha_i\mathbf{u}_i. 
\label{u_m}
\end{equation}
Moreover, the relative velocity between two phases $i$ and $j$ is defined as
\begin{equation}
  \mathbf{u}_{r,ij}  = \mathbf{u}_i - \mathbf{u}_j.
\label{u_r}
\end{equation}
Then, summing the mass conservation equations of all phases and using the relation of Eq. \eqref{eq:closure}, the total mass conservation can be written as
\begin{equation}
\nabla \cdot\mathbf{u}_m=\sum_{i=1}^N \frac{q_i}{\rho_i}-\sum_{i=1}^N\frac{\alpha_i}{\rho_i} \frac{D_i \rho_i}{D t}.
\label{eq:total_mass_conservation}
\end{equation}

Considering the concepts of mixture velocity and relative velocity, it is possible to express the general transport equation for each phase fraction $\alpha_i$ as
\begin{equation}
\begin{aligned}
\frac{\partial \alpha_i}{\partial t}+\nabla \cdot\left(\alpha_i \mathbf{u}_m\right)+\sum_{\substack{j=1 \\
j \neq i}}^N  (\nabla  \cdot  \left.\left(\alpha_i \alpha_j \mathbf{u}_{r, i j}\right)\right)&=\frac{q_i}{\rho_i} + \alpha_i \nabla \cdot\mathbf{u}_m \\
& -\frac{\alpha_i}{\rho_i} \frac{D_i \rho_i}{D t}+\alpha_i \sum_{j=1}^N\frac{\alpha_j}{\rho_j} \frac{D_j \rho_j}{D t}.
\end{aligned}
\label{eq:general_transport}
\end{equation}
Equation \eqref{eq:general_transport} was obtained following the conservative formulation for phase continuity equations developed by \cite{weller2002derivation}, \cite{silva2011development}, and \cite{keser2021eulerian}.

\subsubsection{Capillary effects}

The discontinuity between two moving phases of a multiphase flow on porous media generates an additional relationship between pressure fields denominated as capillary pressure $p_c$, given by:
\begin{equation}
  p_{c,ki} = p^*_k - p^*_i,
\label{eq:capillary_pressure}
\end{equation}
in which $k$ is a reference phase and $i$ is any other phase of the system. 

Using the definition of capillary pressure in Eq. \eqref{eq:capillary_pressure} to replace the pressure of phase $p_i^*$ in Eq. \eqref{eq:modified_momentum}, one can write: 
\begin{equation}
\begin{aligned}
 \dfrac{\partial(\alpha_i\rho_i\mathbf{u}_i)}{\partial t} + \nabla\cdot(\alpha_i \rho_i \mathbf{u}_i\mathbf{u}_i)= -\alpha_i\nabla p^*_k & +\alpha_i\nabla p_{c,ki}  + \alpha_i(\rho_i - \rho_m)\mathbf{g} \\
 &- \alpha_i(\mathbf{g}\cdot \mathbf{h}) \nabla \rho_m  - \alpha_i^2 \frac{\mu_i}{k_{r,i}}\, \mathbf{K}^{-1} \mathbf{u}_i + \mathbf{F}_i.
 \end{aligned}
\label{eq:final_momentum}
\end{equation}

Therefore, the pressure of any phase $i$ is defined as a function of the pressure of the reference phase $k$ and the capillary pressure of the pair $ki$. Note that for the reference phase, the capillary pressure satisfies $p_{c,kk} = 0$, and hence Eq. \eqref{eq:modified_momentum} does not change.

\subsection{Summary of the formulation}

In summary, the mathematical model for compressible multiphase flows in porous media, explained above, consists of a set of partial differential equations, namely the momentum balance equation for each phase fraction, Eq. \eqref{eq:final_momentum}, a total mass balance equation, Eq. \eqref{eq:total_mass_conservation}, and the transport equation of each phase fraction, Eq. \eqref{eq:general_transport}:

\begin{equation}
\begin{aligned}
&  \dfrac{\partial(\alpha_i\rho_i\mathbf{u}_i)}{\partial t} + \nabla\cdot(\alpha_i \rho_i \mathbf{u}_i\mathbf{u}_i)= -\alpha_i\nabla p^*_k  +\alpha_i\nabla p_{c,ki}  + \alpha_i(\rho_i - \rho_m)\mathbf{g} \\
 &\hspace*{4.8cm}- \alpha_i(\mathbf{g}\cdot \mathbf{h}) \nabla \rho_m  - \alpha_i^2 \frac{\mu_i}{k_{r,i}}\, \mathbf{K}^{-1} \mathbf{u}_i + \mathbf{F}_i \\
& \nabla \cdot\mathbf{u}_m=\sum_{i=1}^N \frac{q_i}{\rho_i}-\sum_{i=1}^N\frac{\alpha_i}{\rho_i} \frac{D_i \rho_i}{D t} \\   
&\frac{\partial \alpha_i}{\partial t}+\nabla \cdot\left(\alpha_i \mathbf{u}_m\right)+\sum_{\substack{j=1 \\
j \neq i}}^N  (\nabla  \cdot  \left.\left(\alpha_i \alpha_j \mathbf{u}_{r, i j}\right)\right)=\frac{q_i}{\rho_i} + \alpha_i \nabla \cdot\mathbf{u}_m \\
& \hspace*{7.2cm}-\frac{\alpha_i}{\rho_i} \frac{D_i \rho_i}{D t}+\alpha_i \sum_{j=1}^N\frac{\alpha_j}{\rho_j} \frac{D_j \rho_j}{D t}.
\end{aligned}
\label{eq:complete_system}
\end{equation}

In the next section, we detail the numerical treatment for approximating and solution of the  model system of equations Eq.  \eqref{eq:complete_system}.

\section{Numerical Formulation}%
\label{sec:numerical_formulation}

In this section, we describe the algorithm used to solve the nonlinear system for multiphase flows in porous media along with some models considered and their particularities.

\subsection{Solution approach}
\label{sub:impes}

A segregated scheme based on IMPES \citep{coats2000note} is considered, where the global mass conservation, expressed by a pressure equation, is implicitly solved, while the phase fractions are explicitly solved. 

The numerical model has been implemented in the OpenFOAM framework, that uses the following notation for the momentum balance equation after receiving a finite-volume discretization
\begin{equation}
\mathbf{A}_i\mathbf{u}_i = \mathbf{H}_i(\mathbf{u}_i) - \alpha_i\Big( \nabla p_k^* - \nabla p_{c,ki}  - \left(\rho_i-\rho_m \right)\mathbf{g} + (\mathbf{g}\cdot\mathbf{h})\nabla \rho_m \Big),
\label{eq:OFmomentum}
\end{equation}
where $\mathbf{A}_i$ represents the diagonal term of the matrix system and $\mathbf{H}_i$ represents the off-main diagonal portion of the coefficient matrix with the source term $\mathbf{F}_i$ added.

In order to write a system for pressure and phase fractions,  Eq. \eqref{eq:OFmomentum} is divided by the diagonal coefficient and $\mathbf{u}_i$ is substituted into the total mass conservation Eq. \eqref{eq:total_mass_conservation}, resulting in the following expression:
\begin{equation}
\begin{aligned}
\nabla\cdot \left[\ \sum_{i=1}^{N}\dfrac{\alpha_i}{\mathbf{A}_i}\left(\mathbf{H}_i(\mathbf{u}_i) - \alpha_i\, \Big( \nabla p_k^* - \nabla p_{c,ki}  - \left(\rho_i-\rho_m \right)\mathbf{g} + (\mathbf{g}\cdot\mathbf{h})\nabla \rho_m
\Big)\right)\right] = \\  \sum_{i=1}^N \frac{q_i}{\rho_i} - \sum_{i=1}^{N} \frac{\alpha_i}{\rho_i}\frac{D_i\rho_i}{Dt}.
\end{aligned}
\label{eq:press_total}
\end{equation}
Note that only one pressure equation is obtained for the system instead of a pressure equation for each phase. The total flux $\phi$ resulting from the pressure equation receive the contributions
\begin{equation}
\phi=\phi_p+\phi_{p_c}+\phi_g,
\label{eq:driven_forces}
\end{equation}
where $\phi_p$, $\phi_{p_c}$, and $\phi_g$ represent the flux generated, respectively, by the pressure gradient, capillary pressure and gravity driving forces. The reformulated system is in agreement with the basic framework of OpenFOAM for velocity and pressure coupling. To approximate the pressure-velocity coupling, a solver based on Pressure IMplicit with splitting of operator for Pressure-Linked Equations (PIMPLE) available in OpenFOAM is used  \citep{wang2018literature}.

The phases fractions, in turn, are solved explicitly using the Multidimensional Universal Limiter with Explicit Solution (MULES) \citep{damian2014extended}, an OpenFOAM implementation of the Flux Corrected Transport (FCT) theory \citep{rudman1997volume}.

As a classic finite volume-based numerical scheme, our solver considers the variables calculated in the cell center and interpolated to the faces if necessary. The \textit{upstreamFoam} has an additional option of constructing term by term of the momentum equation interpolated on the faces of the control volumes. This option increases the numerical stability and accuracy of the balance of forces.  To interpolate the different variables distinct numerical schemes are used, for example, linear interpolation, first-order upwind scheme, and harmonic average. More details about the discretization procedures in OpenFOAM can be found in seminal works of \cite{jasak1996error} and \cite{weller1998tensorial}.

\subsubsection{Algorithm}


\begin{figure}
	\tikzstyle{block} = [rectangle, draw,
	text width=23em, text centered, rounded corners, minimum height=2em]
	\tikzstyle{block1} = [rectangle, draw,
	text width=13em, text centered, rounded corners, minimum height=2em]
	\tikzstyle{block2} = [rectangle, draw, fill=gray!20, 
	text width=4em, text centered, rounded corners, minimum height=2em]
        \tikzstyle{noborder} = [rectangle, text width=2em, text centered, minimum height=2em]
	\tikzstyle{line} = [draw, -latex']
	\tikzstyle{line2} = [draw, -]
	\begin{center}	
		\begin{tikzpicture}[node distance = 0.4cm, auto]
		\node [block1] (PC0) {Begin of PIMPLE loop};
		\node [block, below of=PC0, node distance=1.3cm] (PC1) {Solve all the phase fractions};
		\node [block, below of=PC1, node distance=1.45cm] (PC2) {Assemble momentum equation \\for all the moving phases}; 
		\node [block, below of=PC2, node distance=1.55cm] (PC4) {Calculus of the physical effects accounted in the pressure equation}; 
		\node [block, below of=PC4, node distance=1.45cm] (PC5) {Solve the pressure equation};
		\node [block, below of=PC5, node distance=1.4cm] (PC6) {Update the flux of each moving phase and the total flux};
		\node [block, below of=PC6, node distance=1.4cm] (PC7) {Calculate the velocities using the flux};
            \node [block, below of=PC7, node distance=1.4cm] (PC8) {Update of the properties \\ based on pressure and velocity};

		\node [block1, below of=PC8, node distance=1.8cm] (PC9) {iter $<$ nOuterCorrectors \\ \& \\ residual $>$ tol?};
		\node [block2, right of=PC9, node distance=5.0cm] (PC10) {Yes};
		\node[inner sep=0,minimum size=0, node distance=2.0cm,right of=PC10] (PC11) {}; %
		\node [block2, below of=PC9, node distance=1.6cm] (PC12) {No};
		\node [block1, below of=PC12, node distance=1.2cm] (PC13) {End of PIMPLE loop};

            \node [noborder, left of=PC1, node distance=4.3cm] (PC1n) {(1)};
            \node [noborder, left of=PC2, node distance=4.3cm] (PC2n) {(2)};
            \node [noborder, left of=PC4, node distance=4.3cm] (PC4n) {(3)};
            \node [noborder, left of=PC5, node distance=4.3cm] (PC5n) {(4)};
            \node [noborder, left of=PC6, node distance=4.3cm] (PC6n) {(5)};
            \node [noborder, left of=PC7, node distance=4.3cm] (PC7n) {(6)};
            \node [noborder, left of=PC8,node distance=4.3cm] (PC8n) {(7)};
		
		\path [line] (PC0) -- (PC1);
		\path [line] (PC1) -- (PC2);
		\path [line] (PC2) -- (PC4);
		\path [line] (PC4) -- (PC5);
		\path [line] (PC5) -- (PC6);
		\path [line] (PC6) -- (PC7);
		\path [line] (PC7) -- (PC8);
		\path [line] (PC8) -- (PC9);
		\path [line] (PC9) -- (PC10);
		\path [line2] (PC10) -- (PC11);
		\path [line] (PC11) |- (PC1);
		\path [line] (PC9) -- (PC12);
		\path [line] (PC12) -- (PC13);
		
		\end{tikzpicture}
	\end{center}
	\caption{Flowchart of the external loop for one time step.}
	\label{fig:pimpleLoopFlow}
\end{figure}

The numerical scheme applies an external PIMPLE loop at each time step to approximate the problems of ﬂow and transport sequentially. Considering the PIMPLE algorithm \citep{wang2018literature,greenshields2022notes} for pressure-velocity coupling, the flowchart of a PIMPLE loop in \textit{upstreamFoam} is shown in Fig. \ref{fig:pimpleLoopFlow}, detailing the main operations. When the loop begins, all the phase fractions are solved (step 1). The MULES algorithm is used for the moving phases, while a diagonal solver is used for the stationary phases due to the simplicity of the mass balance when the phase is stationary.

Next, the momentum equation is assembled for all moving phases (step 2), noting that there is no momentum prediction in the \textit{upstreamFoam} formulation. Since the phase fractions at faces are repeatedly used in subsequent steps, they are interpolated and stored in a new field for efficiency. The physical effects that must be accounted for in the pressure equation, such as gravitational effects, capillary pressure, compressibility, and dilatation, are calculated in step 3. Finally, the pressure equation is solved (step 4).

With the phase fractions and pressure solved, the flux of each moving phase and the total flux are updated (step 5). The velocities at the cell centers are then obtained (step 6) using the fluxes calculated in the previous step. Finally, with updated fields of phase fraction, pressure, and velocities, other properties based on these fields such that relative permeability, density and viscosity are updated. These seven steps are repeated until the tolerance, that is based on a residual related to the change in pressure solution, or the maximum number of iterations (nOuterCorrectors) is reached. We remark that if only one external iteration is defined, the algorithm falls back into a scheme similar to IMPES.

\subsubsection{Time-step limitations}

Due to the explicit treatment, the time steps need to satisfy the Courant-Friedrichs-Lewy (CFL) condition \citep{courant1928partiellen}. Since there is no single optimal criterion always ensuring stability and efficiency \citep{franc2016benchmark}, the \textit{upstreamFoam} has two classic criteria for the time step choice: the Courant condition \citep{courant1928partiellen}, and the Coats restriction \citep{coats2003impes}. In the implemented models, the time step is adaptively adjusted such that the following limit is not exceeded:
\begin{equation}
	\Delta t_{\max} = \frac{C_{\max}}{C},
	\label{eq:delta_t_max}
\end{equation}
where $C_{\max}$ is a user-defined limit and $C$ is a restriction defined by each stability criterion according to the flow properties.

The classic Courant number condition is given by
\begin{equation}
	C = C_{\text{Courant}} =  \frac{1}{2}\max_{\alpha_i,\, j} \left( \frac{ \sum_{f=1}^{m_j} |\phi_{f}|}{V_j} \right) \Delta t,
	\label{eq:Co_condition}
\end{equation}
where $\phi_{f}$ are the total fluxes of phase fraction $\alpha_i$ through each face $f$ of cell $j$, $m_j$ is the total number of faces of cell $j$, $V_j$ is the volume of cell $j$, and $\Delta t$ is the previous time step.

The Coats criterion \citep{coats2003impes} is also considered by the \textit{upstreamFoam}. We use a system containing a wetting phase $w$ and a non-wetting phase $n$ to describe this condition, which in this case is given by
\begin{equation}
	C = C_{\text{Coats}}=  \max_{j} \left[\frac{1}{\alpha_v\,V_j}\left(\Psi\frac{\partial p_c}{\partial S_w}\,\sum_{f=1}^{m_j}T_f+\frac{\partial F_w}{\partial S_w}\,\sum_{f=1}^{m_j} \phi_{f} \right)\right] \Delta t,
\label{eq:Coats_condition}
\end{equation}
where $T_f$ is the transmissibility of face $f$ of cell $j$
\begin{equation}
T_f=\frac{K_f A_f}{\Delta x_f},
\end{equation}
$A_f$ is the area of face $f$, $\Delta x_f$ is the distance between the centers of two neighboring cells that contain face $f$, and $K_f$ is the harmonic interpolation of absolute permeability at face $f$. Defining the phase mobilities $\lambda_i=k_{r,i}/\mu_i$, $i\in\{w,n\}$, one have that $\Psi$ is the harmonic average of mobilities
\begin{equation}
\Psi=\frac{2\lambda_w\lambda_n}{\lambda_w+\lambda_n}
\end{equation}
and $F_w$ is the fractional flow
\begin{equation}
F_w=\frac{\lambda_w}{\lambda_w+\lambda_n}.
\end{equation}
In Eq. (\ref{eq:Coats_condition}), the evaluated $\phi_f$ considers the sum of the total fluxes of both phases through face $f$. \cite{coats2003impes} shows a formulation of the Coats criterion for a three-phase system.

Note that the Courant condition depends on fluxes of phases, while the Coats restriction considers the same fluxes and additional information related to the fractional flow and capillary pressure effects. Therefore, the Coats stability criterion is more adapted to the porous media problems, however, it may be very restrictive in certain circumstances and is not necessarily the optimum choice \citep{franc2016benchmark}. As both criteria are able to avoid numerical instabilities due to the non-linear effects involved in the problems considered in this paper, we control the time step variations by Coats restriction for the cases with capillary pressure and by Courant condition for the remaining cases. We refer to the comparative study by \cite{franc2016benchmark} for more detailed discussions about the effectiveness of each criterion at different porous media flow regimes.

\subsection{Porous media modeling}
\label{sub:porous_models}

In this section, we describe the models used to deal with the essential features of multiphase flows in porous media.

\subsubsection{Relative permeability models}

A generic framework for multiphase flows with any number of phases requires relative permeability models with respect to each ﬂuid. We apply the widely used Brooks and Corey model \citep{brooks1965hydraulic}, which relates the relative permeability of each phase to the phase fraction by
\begin{equation}
k_{r,i}=k_{r,i (\max)}\left(\frac{\alpha_i}{\alpha_v}\right)^n,
\label{eq:Krel_BC}
\end{equation}
where $n$ is a power coefficient associated to the porous media properties, and $k_{r,i (\max)}$ is the maximal relative permeability.

Relative permeability functions depend on the particular porous medium and ﬂuid phases, and they are obtained by performing displacement experiments on samples that represent the real conditions of each porous medium, if possible. The same applies to the exponent of the Brooks and Corey model, which is usually obtained by fitting experimental data. However, there are some standard choices, for example, quadratic functions, whose application is adequate to represent the non-linear coupling between flow and transport in porous media \citep{aziz1979petroleum}. It is worth noting that it is also possible to find equivalence between the parameters of the Brooks and Corey model to other models in the literature \citep{morel1996parameter}.

The \textit{upstreamFoam} also allows for the use of tabulated methodology, considering a user provided table with the relative permeability value as a function of the saturation.

\subsubsection{Capillary pressure models}

We consider classical approximations that $p_c$ depends only on saturation, and hence
\begin{equation}
\nabla p_c=p'_c\ \nabla\Big(\frac{\alpha_i}{\alpha_v}\Big).
\label{eq:pc_BC_grad}
\end{equation}

A well-known correlation considered here for the capillary pressure is the Brooks and Corey model \citep{brooks1965hydraulic}, which is given by the following expression 
\begin{equation}
p_c=p_{c,0}\,\left(\frac{\alpha_i}{\alpha_v}\right)^{-\beta},
\label{eq:pc_BC}
\end{equation}
where $p_{c,0}$ is the entry capillary pressure and $\beta$ is a parameter related to the pore size distribution. 

An application of a user-defined table with capillary pressure derivative value as a function of the saturation is also possible. Another possibility is the computation of the capillary pressure through interpolations performed with tabulated data of the Leverett J-function \citep{leverett1941capillary} provided by the user.

\subsubsection{Compressibility models}

To consider compressible phases, we assume that the density can be expressed in terms of the pressure through compressibility terms. For example, in the case of a compressible formation, the porosity, that is, the void fraction $\alpha_v$, is a pressure dependent unknown.

A basic model considered for compressible flows computes the density variation according to a constant compressibility, such that: 
\begin{equation}
\rho = \rho_{0}\, e^{c(p-p_{0})},
\label{eq:rhoConst}
\end{equation}
where $c$ is the compressibility of the phase, $\rho_{0}$ is a reference density, and $p_{0}$ is a reference pressure. It is derived from an equation of state and can be applied to most liquids that do not contain large amounts of dissolved gas \citep{chen2006computational}. If the flow is slightly compressible, which corresponds to the cases considered in our experiments, it is enough to use the following linear approximation:
\begin{equation}
\rho = \rho_{0}\, (1+c(p-p_{0})).
\label{eq:rhoLinear}
\end{equation}
For gas flow, gas compressibility is generally not considered constant, and a form that considers the gas law is required \citep{aziz1979petroleum}.

Other possibility available in \textit{upstreamFoam} is the application of a user-defined table with density value as a function of the pressure. We remark that other models to simulate compressible moving phases implemented on the thermophysical classes of OpenFOAM \citep{nguyen2022real} can be easily incorporated into the \textit{upstreamFoam}.

\section{Numerical Results}
\label{sec:numerical_results}

In this section, the multiphase model for flows in porous media is used in several experiments for verification and application in realistic data. Initially, we study classical problems by comparing the results obtained with analytical or semi-analytical solutions. Then, the solver is applied to approximate reservoir scale problems, including the realistic SPE10 benchmark \citep{christie2001tenth}. To close, a multiphase flow simulation in a heterogeneous core is presented.

The main numerical schemes used in our experiments include explicit Euler time integration, linear discretizations for gradients, and the first-order upwind method for divergents. The linear system is solved with stabilized preconditioned conjugate gradient (PCG) method combined with a generalized geometric algebraic multigrid (GAMG) method and diagonal incomplete Cholesky (DIC) smoother considering a tolerance of $10^{-8}$.  The PIMPLE loop is defined with a tolerance of $10^{-4}$ and 10 outer iterations.

Extensive research has demonstrated significant improvements in the scalability and parallel performance of OpenFOAM in high-performance computing (HPC) environments. For instance, studies by \cite{galeazzo2024performance} have provided in-depth analyses of OpenFOAM's performance on various HPC architectures, highlighting its capability to efficiently utilize modern computing resources. Given that \textit{upstreamFoam} is fully based on OpenFOAM, its parallel performance is expected to align closely with that of the underlying framework. This implies that any advancements in OpenFOAM’s parallel efficiency will likely result in corresponding improvements in \textit{upstreamFoam}. The speed-up results showed by \cite{horgue2015open} for an OpenFOAM-based toolbox for porous media confirm this behavior as it is observed a super linear speed-up in a certain range, aligned with the observation in the work of \cite{galeazzo2024understanding}.

As the primary objective of the current work is to present the formulation and validation of the implemented code, all the simulations in the following sections were executed using just one core (serial) of an Intel Core i7-12700H processor CPU 4.7GHz with 64GB of memory.

\subsection{Verification cases}
\label{sub:verification}

We verify the solver using classical problems with analytical or semi-analytical solutions. Four cases are considered, the first and second ones are homogeneous and heterogeneous Buckley-Leverett problems, the third one is a capillary-gravity equilibrium, and the last one is a compressible formation case. 
The validation cases were chosen such that relative permeability, capillary-gravity forces, and compressibility effects can be evaluated separately. The objective is to verify whether the new solver produces results that converge to the reference solutions when applied to different situations since it will then be used to simulate realistic problems with all these effects.

\subsubsection{Homogeneous Buckley-Leverett}

The behavior of a two-phase immiscible and incompressible flow in one-dimensional porous medium can be investigated using the Buckley-Leverett semi-analytical solution \citep{wu2015multiphase}. 

In this study, we perform a numerical verification considering a channel of 0.065\,m long saturated with oil (denoted by the subscript $o$), where water (denoted by the subscript $w$) is injected from left at a constant volumetric rate of $1.66\times 10^{-8}\,\mbox{m/s}$, and the pressure at right is fixed in 0.1\, MPa. This case considers homogeneous porosity of $\alpha_v=0.2$ and absolute permeability of $\mathbf{K}=1\times 10^{-13}\,\mbox{m}^2$. The gravity and capillary pressure are negligible. 
The phase densities are $\rho_w=1000\,\mbox{kg/m}^{3}$ and $\rho_o=800\,\mbox{kg/m}^{3}$, while the viscosities are $\mu_w=0.001\, \mbox{Pa}\cdot\mbox{s}$ and $\mu_o=0.002\, \mbox{Pa}\cdot\mbox{s}$. The Brooks and Corey relative permeability model with $n=2$ and $k_{r,i(\max)}=1$ is applied for both phases. The Courant time step control is selected with $C_{\max}=0.5$.

\begin{figure}[htb!]
  \centering
  \includegraphics[width=7cm]{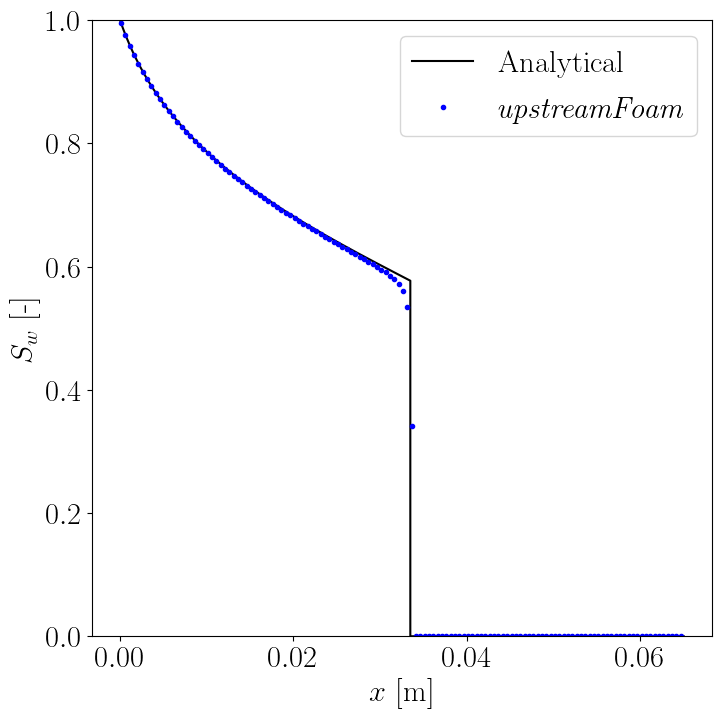}
  \caption{Saturation profile after 1800\,s of the homogeneous Buckley-Leverett case.}
  \label{fig:homogeneous_BL}
\end{figure}

In Fig. \ref{fig:homogeneous_BL} the saturation profile after 1800\,s is shown for a computational mesh with 500 cells, where it is possible to note that the approximation is close to the analytical solution. 
A convergence study is reported in Table \ref{tab:BL_homo}, where $L^1$ errors are compared for different meshes. We observe errors decreasing substantially in a linear behavior, that is the expected slope due to the first-order discretizations. Additionally, the execution times are shown. 

\begin{table}
\centering
\begin{tabular}{|c|c|c|}
\hline 
Number of cells [-] & $L^1$ error [-] & Execution time [s] \\ \hline 
500          & 0.003637     & 1.21    \\ \hline 
1000         & 0.001968     & 3.14    \\ \hline 
2000         & 0.001022     & 9.26    \\ \hline 
4000         & 0.000370     & 31.35   \\ \hline 
\end{tabular}
\caption{$L^1$ errors and execution times for the homogeneous Buckley-Leverett case.}
\label{tab:BL_homo}
\end{table}

\subsubsection{Heterogeneous Buckley-Leverett}

An extension of the previous analysis to flows in one-dimensional heterogeneous porous medium has been introduced by \cite{wu2015multiphase}, in which the formation consists of a number of domains with different rock properties.

In order to consider the heterogeneous case, we assume that the porosity is $\alpha_v=0.2$ if $0\leq x \leq 0.0325$, and $\alpha_v=0.1$ if $0.0325<x \leq 0.065$. In the Brooks and Corey relative permeability model, we consider $n=1$ for both phases if $x\leq0.325$ and $n=4$ for both phases if $x>0.325$. All the other parameters consider the same setup of the previous example.

\begin{figure}[htb!]
  \centering
  \includegraphics[width=7cm]{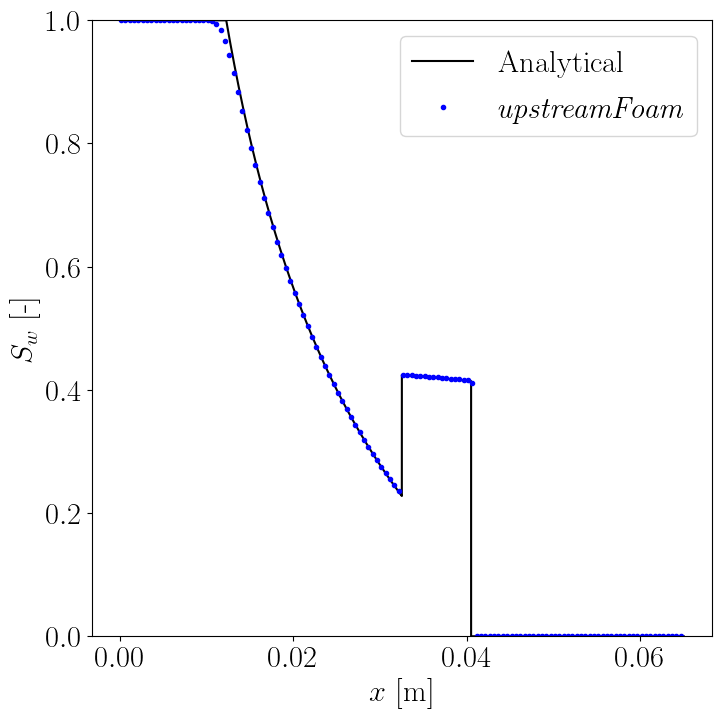}
  \caption{Saturation profile after 1800\,s of the heterogeneous Buckley-Leverett case.}
  \label{fig:heterogeneous_BL}
\end{figure}

In Fig. \ref{fig:heterogeneous_BL} the saturation profile after 1800\,s is shown for a computational mesh with 500 cells, where we also note that the approximation is close to the analytical solution for the heterogeneous case. 
Table \ref{tab:BL_het} shows the respective convergence study, with errors decreasing as the mesh is refined. 
The behavior of the execution time when refining the mesh is also shown, where we can note an increase in the computational cost in relation to the homogeneous case.
\begin{table}
\centering
\begin{tabular}{|c|c|c|}
\hline 
Number of cells [-] & $L^1$ error [-] & Execution time [s] \\ \hline 
500          & 0.004342     & 1.96   \\ \hline 
1000         & 0.002499     & 5.25   \\ \hline 
2000         & 0.001416     & 16.40  \\ \hline 
4000         & 0.000889    & 58.84   \\ \hline 
\end{tabular}
\caption{$L^1$ errors and execution times for the heterogeneous Buckley-Leverett case.}
\label{tab:BL_het}
\end{table}

\subsubsection{Capillary-gravity equilibrium}

We now study an air-water incompressible flow in a vertical column (1\,m tall) with capillary and gravity effects. The flow is carried out until the gravity-capillarity equilibrium, starting with an initial distribution of the water saturation in a step-wise fashion: the lower half is partially saturated with water ($S_w=0.5$ and $S_a=0.5$), while the upper half is dry ($S_w=0$ and $S_a=1$). 
The bottom boundary condition is a Neumann zero condition, and the top boundary condition is a Dirichlet fixed pressure of 0.1\,MPa. 

Homogeneous porosity of $\alpha_v=0.2$ and absolute permeability of $\mathbf{K}=1\times 10^{-11}\,\mbox{m}^2$ are considered. The phase densities are $\rho_w=1000\,\mbox{kg/m}^{3}$ and $\rho_a=1\,\mbox{kg/m}^{3}$, while the viscosities are $\mu_w=0.001\, \mbox{Pa}\cdot\mbox{s}$ and $\mu_a=1.76\times 10^{-5}\, \mbox{Pa}\cdot\mbox{s}$. The Brooks and Corey relative permeability model with $n=1$ and $k_{r,i(\max)}=1$ is applied for both phases. Moreover, the capillary pressure is defined through a table filled by the Brooks and Corey correlation using 501 points with parameters $p_{c,0}=1000$\,Pa and $\beta=0.5$. The Coats criteria has been applied, with $C_{\max}=0.5$.

Following the analysis proposed by \cite{horgue2015open} and \cite{carrillo2020multiphase}, the theoretical steady state can be described as the balance between capillary and gravitational forces, described by 
\begin{equation}
\frac{\partial p_c}{\partial y}=(\rho_a-\rho_w)g_y,
\label{eq:derivada_pc_equilibrio}
\end{equation}
where $g_y=-9.8\,\mbox{m/s}^{2}$ is the gravity in the main flow direction. The expression above can be written as
\begin{equation}
\frac{\partial S_w}{\partial y}=\frac{(\rho_a-\rho_w)g_y}{\frac{\partial p_c}{\partial S_w}},
\label{eq:derivada_sat_equilibrio}
\end{equation}
which allows for the calculation of the water saturation gradient according to the chosen capillary pressure model.

\begin{figure}[htb!]
  \centering
  \subfigure[Saturation profile \label{fig:homogeneous_pc_gravity_a}]
  {\includegraphics[scale=0.35]{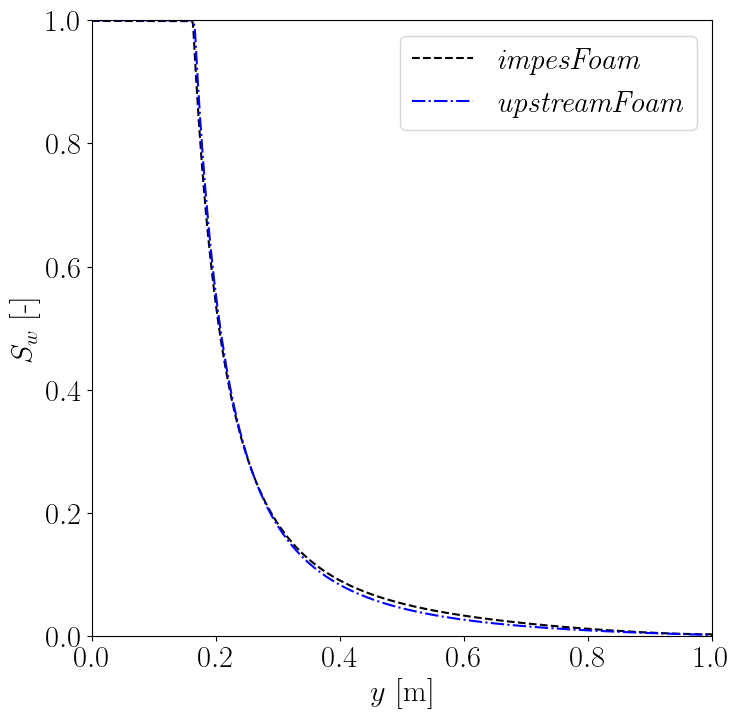}}
  \subfigure[Saturation gradient \label{fig:homogeneous_pc_gravity_b}]
  {\includegraphics[scale=0.35]{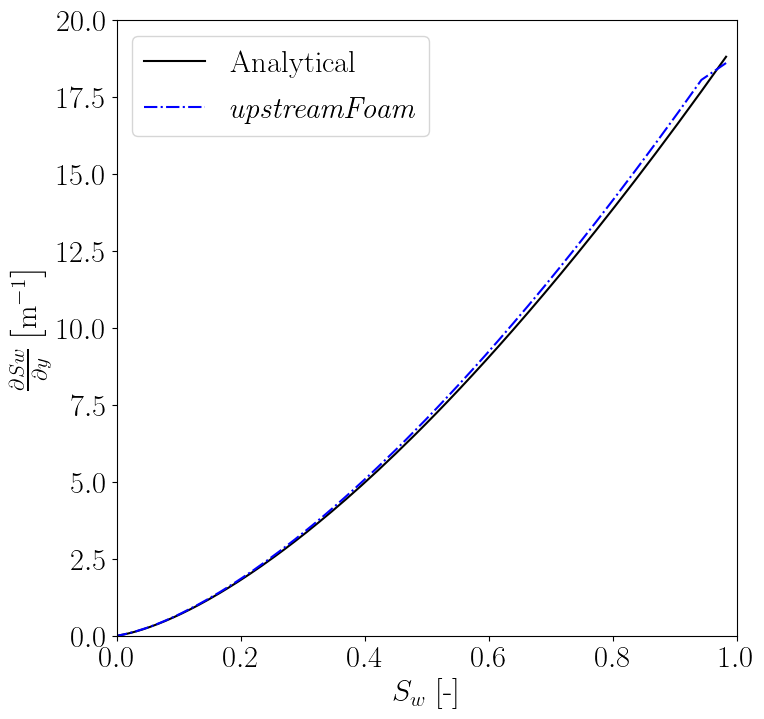}}
  \caption{Capillary-gravity equilibrium test.}
\end{figure}

The saturation profile at the capillary-gravity equilibrium (after 20000\,s) for a computational mesh with 300 cells can be seen in Fig. \ref{fig:homogeneous_pc_gravity_a}. We remark that this case was presented by \cite{horgue2015open} for the development of the \textit{porousMultiphaseFoam} toolbox. Therefore, a comparison between both approximations with equivalent numerical setup is included, specifically using the solver \textit{impesFoam} available in \textit{porousMultiphaseFoam}. It is possible to note that the obtained saturation profiles are related. A comparison between theoretical and numerical evaluations of the water saturation gradient can be seen in Fig. \ref{fig:homogeneous_pc_gravity_b}, in which we see that both solutions are very similar.

In Table \ref{tab:CG_equilibrium}, the convergence study along with computational times for the capillary-gravity test are presented. It is possible to note that the errors decrease as the mesh is refined, while the computational time is increased. 
Note that, due to the effect of capillary pressure, the computational cost for this experiment increased significantly compared to the previous examples. 
Furthermore, the \textit{impesFoam} required a computational time of 401.01\,s for the mesh with 300 cells, which corresponds to less than half the computational time required by the \textit{upstreamFoam}. However, when comparing the computational time between \textit{upstreamFoam} and \textit{impesFoam} we need to keep in mind that the latter is a solver for two-phase flows with only constant porosity fields and without considering compressibility effects. In other words, the \textit{upstreamFoam} proposal is not to be computationally faster than \textit{impesFoam}, but rather to add the modeling of physical complexities that enable a greater variety of applications. 
\begin{table}
\centering
\begin{tabular}{|c|c|c|}
\hline 
Number of cells [-] & $L^1$ error [-] & Execution time [s] \\ \hline 
30          & 0.2413     & 7.88    \\ \hline 
75          & 0.09575     & 48.52   \\ \hline 
150         & 0.048429     & 227.1  \\ \hline 
300         & 0.042366     & 1000.86   \\ \hline 
\end{tabular}
\caption{$L^1$ errors and execution times for the capillary-gravity equilibrium test.}
\label{tab:CG_equilibrium}
\end{table}

\subsubsection{Compressible formation}

Since the formation compressibility can significantly influence the reservoir estimations, we propose a simplified  verification of the total mass of oil produced in a scenario of depletion generated by compressibility effects. The present experiment considers an oil single-phase flow, however, the \textit{upstreamFoam} framework performs the computation of both porous and fluid phases.

We study the flow of an incompressible oil in a one-dimensional compressible reservoir of 1000\,m long with 100 computational cells. The reservoir, filled with oil, is initially under a pressure of 100\,MPa, and a fixed pressure of 10\,MPa is maintained at the outlet, resulting in a transient state of oil production by system depletion. 
The initial porosity, that is, the initial void fraction of the system, is homogeneous, with a value of $\alpha_v=0.3$. 
We consider a constant compressibility of $c_v=1\times 10^{-9}\,\mbox{Pa}^{-1}$, absolute permeability of $\mathbf{K}=1\times 10^{-12}\,\mbox{m}^2$, oil density of $\rho_o=800\,\mbox{kg/m}^{3}$, and oil viscosity of $\mu_o=0.002\, \mbox{Pa}\cdot\mbox{s}$. In this experiment, gravity effects are neglected. 

Once the porous medium is compressible, the void fraction varies with pressure, and can be described by using a compressibility model, as for example,
\begin{equation}
\alpha_v = \alpha_{v,0}\, (1+c_v(p_v-p_{v,0})),
\label{eq:porosity_rhoLinear}
\end{equation}
where $c_v$ is the compressibility of the formation, $\alpha_{v,0}$ is the reference porosity, and $p_{v,0}$ is the reference pressure. The above correlation allows for calculating a semi-analytical estimation for the mass balance between the initial and final states of the system.

Considering that the mass of the formation is conserved throughout the depletion process, the initial oil mass is $m^i_{o}=\alpha^i_{v}\,\rho_{o}\,V$, while the final oil mass is $m^f_{o}=\alpha^f_{v}\,\rho_{o}\,V$. Therefore, the total mass of oil produced, $m^p_{o}$, can be calculated by
\begin{equation}
m^p_{o}=m^i_{o}-m^f_{o}=(\alpha^i_{v} - \alpha^f_{v})\,\rho_{o}\,V.
\label{eq:oil_total_mass}
\end{equation}
Using the linear model of Eq. \eqref{eq:porosity_rhoLinear} in Eq. \eqref{eq:oil_total_mass}, gives the following expression for the total mass of oil produced:
\begin{equation}
m^p_{o}=\alpha_{v,0}\, c_v(p^i_v-p^f_v)\,\rho_{o}\,V.
\label{eq:oil_total_mass_final}
\end{equation}

For the simulation parameters considered above, an oil production of $5.4\times 10^{7}$\,kg, i.e., 424\,562.18\,STB after depletion is estimated. To evaluate the amount of produced mass of oil during the simulation, we test different values of time step size. The results are reported in Table \ref{tab:mass_oil}, where it is possible to note that the solver produces adequate results with the error decreasing as the time step decreases. 


\begin{table}
\centering
\begin{tabular}{|c|c|c|}
\hline 
Time step [s] & Oil produced [STB] & Error [\%] \\
\hline 
2400          &411196.33     & 3.15       \\ \hline 
1200          &415127.46     & 2.22       \\ \hline 
600           &418272.36     & 1.56       \\ \hline 
300           &419844.82     & 1.09       \\ \hline 
150           &421417.27     & 0.75       \\ \hline 
\end{tabular}
\caption{Mass of oil produced according to the time step size.}
\label{tab:mass_oil}
\end{table}

\begin{figure}[htb!]
  \centering
  \includegraphics[width=12cm]{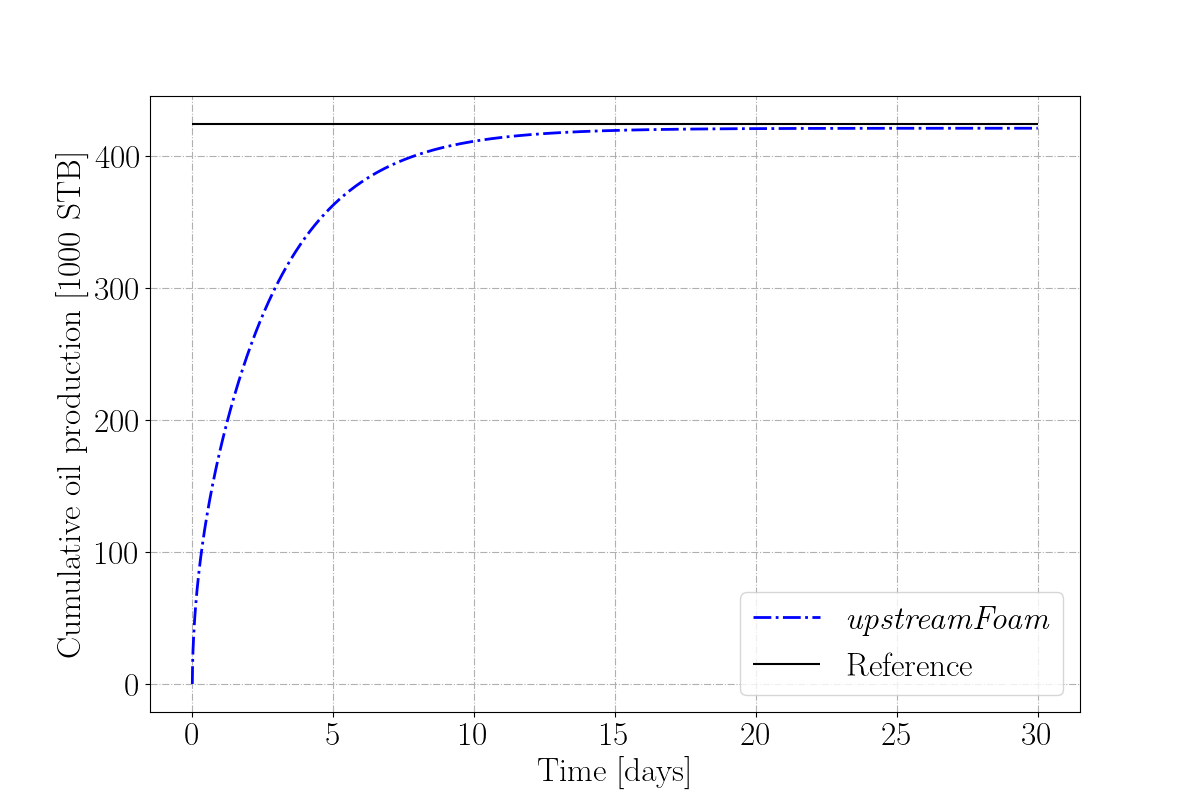}
  \caption{Mass of oil produced by the \textit{upstreamFoam} with $\Delta t =150$\,s.}
  \label{fig:mass_oil}
\end{figure}

The graph in Fig. \ref{fig:mass_oil} illustrates the curve of oil produced for $\Delta t =150$\,s. Note that the calculated mass of oil produced approaches the expected value.

\subsection{2D reservoir problems}
\label{sub:reservoir}

This section presents the application of \textit{upstreamFoam} to approximate two reservoir scale problems. The first experiment shows the evaluation of the head loss of an oil flow in a circular reservoir. The second experiment presents a gas-oil flow in a heterogeneous reservoir with data given by the realistic SPE10 benchmark \citep{christie2001tenth}.

\subsubsection{Oil flow in a circular reservoir}

The head loss of the reservoir is a typical estimate in oil recovery studies. In this experiment, the head loss of a 2D reservoir with oil production by system depletion is investigated. 

We simulate the flow of an incompressible oil in a compressible reservoir, given by a circular geometry. The reservoir is initially fully saturated with oil, which is produced through a sink term located in the center of the domain. A fixed pressure of $19.61$\,MPa is maintained at the boundaries, along with a constant flow rate of $q=0.36\times 10^{-4}\mbox{m}^3/\mbox{s}$ at the sink cell. The reservoir, with a diameter of 1000\,m, contains an initial homogeneous void fraction of $\alpha_v=0.3$. Linear compressibility model is considered with $c_v=1\times 10^{-9}\,\mbox{Pa}^{-1}$, absolute permeability of $\mathbf{K}=2\times 10^{-13}\,\mbox{m}^2$, oil density of $\rho_o=880\,\mbox{kg/m}^{3}$, and oil viscosity of $\mu_o=0.002\, \mbox{Pa}\cdot\mbox{s}$. In this experiment, gravity effects are neglected. The Courant time step control is selected with $C_{\max}=0.5$.

\begin{figure}[htb!]
  \centering
  \includegraphics[width=10cm]{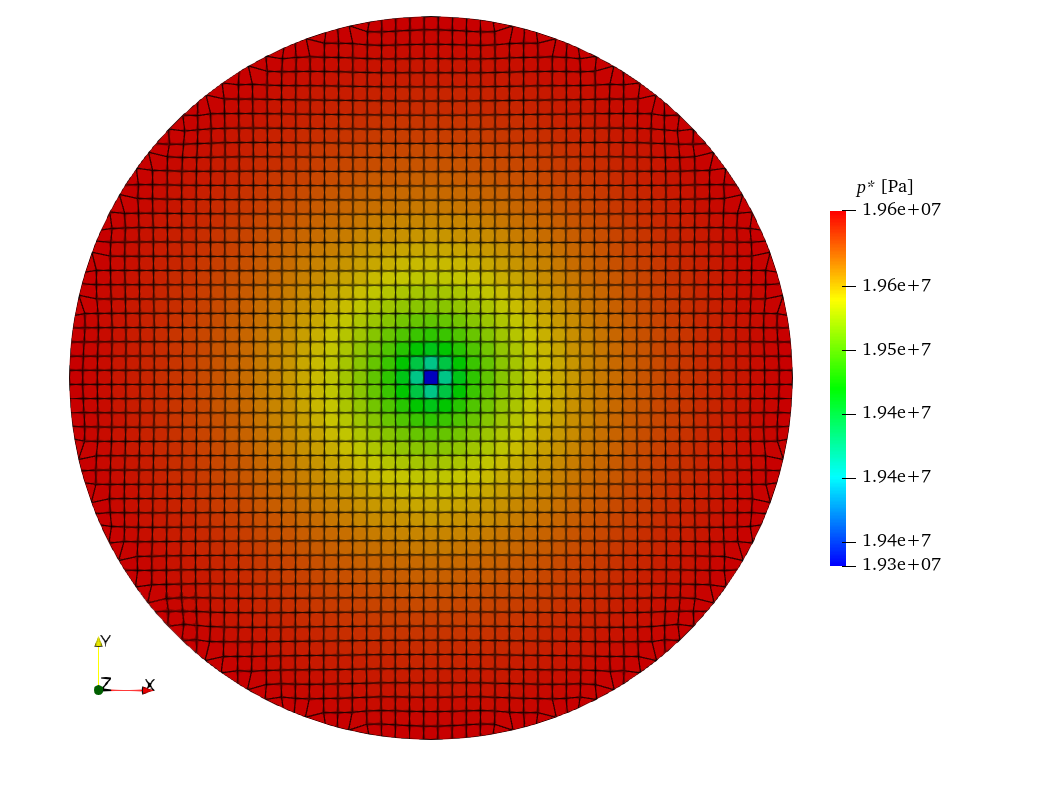}
  \caption{Pressure profile of the 2D reservoir with a sink cell in the center of the domain.}
  \label{fig:head_loss}
\end{figure}

Figure \ref{fig:head_loss} shows the pressure profile at time $t=182.64$ days yielded by a simulation with 2053 computational cells. In this case, the cell size is $\Delta x=\Delta y= 19.6$\,m. From the pressure profile, it is noted that the flow is symmetrical towards the producing cell.

\begin{figure}[htb!]
  \centering
  \includegraphics[width=10cm]{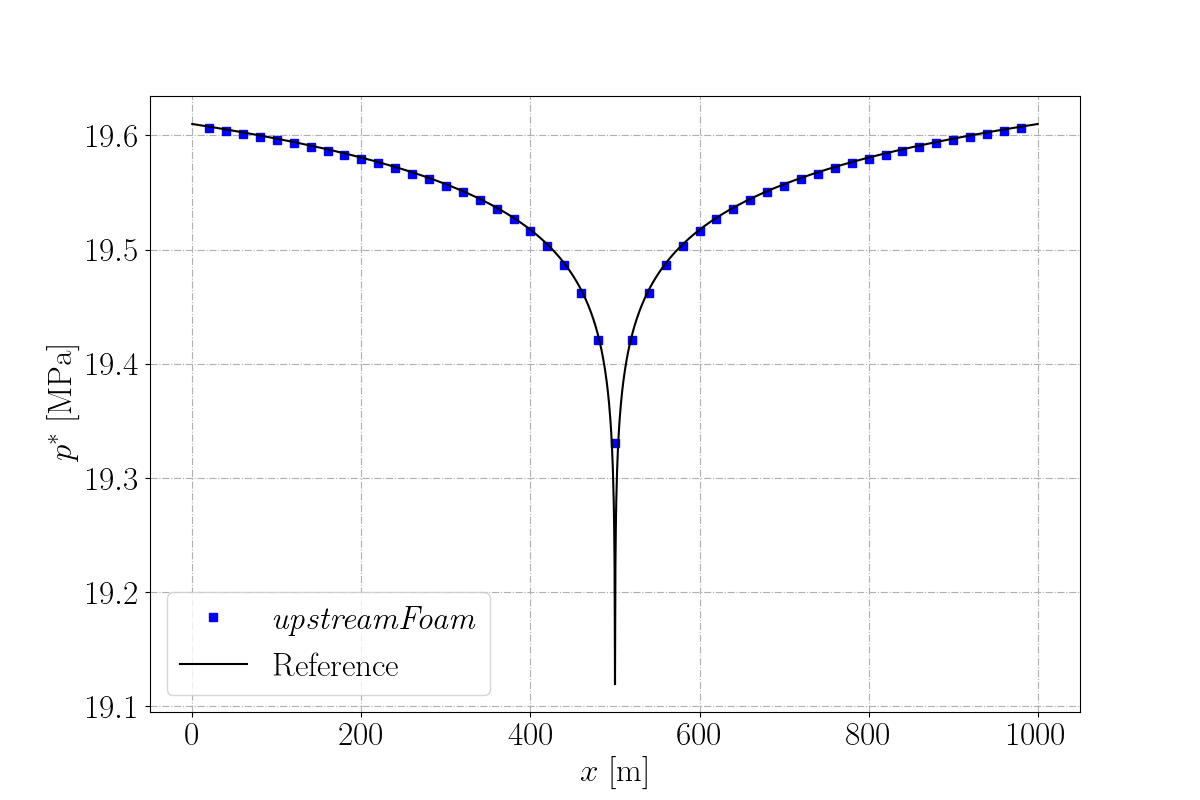}
  \caption{Pressure distribution inside the 2D reservoir as a function of the diameter.}
  \label{fig:head_loss_line}
\end{figure}

Considering a radial reservoir and using mass conservation and Darcy law for a single-phase flow, one can write 
\begin{equation}
\frac{dp}{dr}=-\frac{\mu\, \mathbf{K}^{-1} q}{2\pi \rho rh},
\label{eq:head_loss_radial}
\end{equation}
where $r$ is the radius of the reservoir, $q$ is the mass production, $h$ is the height of the sink cell, and $\rho$ is the fluid density \citep{chen2009well}. After integrating from the wellbore $r_w$ to the reservoir radius $r$, we obtain:
\begin{equation}
\Delta p=-\frac{\mu\, \mathbf{K}^{-1} q}{2\pi \rho h}\ln\left(\frac{r}{r_w}\right),
\label{eq:head_loss_Peaceman}
\end{equation}
which coincides to the Peaceman well model considering isotropic permeabilities, square grid, single-phase flow, and the well located at the center of an interior cell \citep{peaceman1978interpretation}. Therefore, we are able to estimate the pressure drop between any location in the reservoir and the sink cell by Eq. \eqref{eq:head_loss_Peaceman}. Figure \ref{fig:head_loss_line} shows the pressure distribution inside the reservoir presented in Fig. \ref{fig:head_loss} as a function of the diameter, where we can note that the pressure approximation for the reservoir cells matches the reference solution (real reservoir pressure), while the central cell presents the greatest contribution to the discrepancies due to the well modeling approximation.  

Once a fixed pressure is applied at the boundaries, the numerical head loss of the system can be computed by using the pressure value in the producer cell. 
In order to obtain a theoretical head loss of 0.49\,MPa, we set in Eq. \eqref{eq:head_loss_Peaceman} the values of $h=1$\,m and $r_w = 0.1$\,m. We remark that in this experiment a value of $r=500$\,m is considered. Table \ref{tab:head_loss} reports the average error between the numerical pressure obtained by the \textit{upstreamFoam} and the reference at time $t=182.64$ days considering different grid sizes, with $\Delta x=\Delta y$. The results show that with the proposed solver it is possible to obtain accurate estimates of head loss with low errors when compared to the theoretical values.

To indicate the potential of the code, the computational efficiency in this two-dimensional case was calculated using the following estimate
\begin{equation}
    e_j = \dfrac{n_{j}/n_{j-1}}{t_{j}/t_{j-1}}\, 100\%,
\end{equation}
where $n_j$ indicates the number of cells and $t_j$ the execution time for the $j$-th mesh. 
The execution times lead to a computational average efficiency of 122.03\%, that represents an excellent scalability rate. 
The efficiency evaluation is not present for the one-dimensional cases because 1D configurations are less suitable for evaluating the performance of the solver.
\begin{table}
\centering
  \begin{tabular}{|c|c|c|c|c|}
    \hline 
    Cells [-] & $\Delta x$ [m] & Error [\%] & Execution time [s] & Efficiency [\%] \\
    \hline 
	489    & 40.7 & 0.077 & 0.63 & - \\ \hline 
	2053   & 19.6 & 0.025 & 1.51 & 175.16 \\ \hline 
	8021   & 9.90 & 0.008 & 4.7 & 125.52 \\ \hline 
	31757  & 4.08 & 0.005 & 19.97 & 93.18 \\ \hline 
	126309 & 2.49 & 0.003 & 84.26 & 94.26 \\ \hline 
  \end{tabular}
\caption{Average errors and execution times according to the grid size for the incompressible case.}
\label{tab:head_loss}
\end{table}

Lastly, for the same scenario, we estimate the numerical head loss of a compressible oil flow, whose density is given by
\begin{equation}
\rho=\frac{\rho_{o,s} + R_s\, \rho_{g,s}}{B_{ob}\Big(1-c_{o}(p-p_b)\Big)},
\label{eq:density_oil_compress}
\end{equation}  
where $\rho_{o,s}$ and $\rho_{g,s}$ are the densities of oil and gas at standard conditions, $R_s$ is the gas solubility in oil, $B_{ob}$ is the formation volume factor at the bubble point pressure $p_b$, and $c_o$ is the oil compressibility. Results at time $t=182.64$ days for $\rho_{o,s}=880\,\mbox{kg/m}^{3}$, $\rho_{g,s}=1\,\mbox{kg/m}^{3}$, $R_s=199.0$, $B_{ob}=1.4887$, $p_b=14.71$\,MPa, and $c_o=1.77\times 10^{-9}\,\mbox{Pa}^{-1}$ are shown in Table \ref{tab:head_loss_compress}. We can observe accurate estimates for the numerical pressure, with the errors decreasing as the mesh is refined in the same behavior that the obtained in the incompressible case. The computational time is increased in a proportion such that an average efficiency of 128.26\% is attained.
\begin{table}
  \centering
  \begin{tabular}{|c|c|c|c|c|}
    \hline 
    Cells [-] & $\Delta x$ [m] & Error [\%] & Execution time [s] & Efficiency [\%] \\
    \hline 
	489    & 40.7 & 0.077 & 0.71 & - \\ \hline 
	2053   & 19.6 & 0.026 & 1.56 & 187.82 \\ \hline 
	8021   & 9.90 & 0.008 & 5.05 & 116.35 \\ \hline 
	31757  & 4.08 & 0.006 & 21.13 & 102.22 \\ \hline 
	126309 & 2.49 & 0.003 & 88.39 & 106.65 \\ \hline 
  \end{tabular}
  \caption{Average errors and execution times according to the grid size for the compressible case.}
  \label{tab:head_loss_compress}
\end{table}

\subsubsection{Gas-oil flow in a heterogeneous reservoir}

Petroleum reservoirs present high contrast in many proprieties, which impacts the accuracy of the numerical solution and production estimates. This experiment solves a high-contrast heterogeneous reservoir problem with realistic data provided by the SPE10 benchmark \citep{christie2001tenth}.  

\begin{figure}[htb!]
  \centering
  \includegraphics[width=13cm]{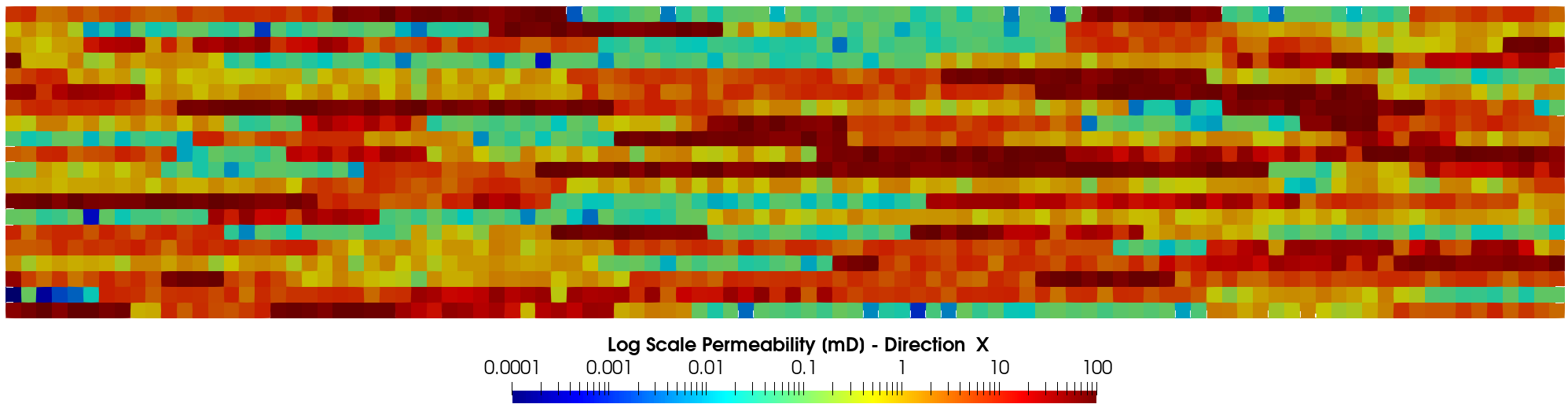}
  \caption{Log-scaled permeability field of the SPE10 project.}
  \label{fig:perm_SPE10}
\end{figure}

We select the first model, which consists of a two-phase flow (gas and oil) in a 2D vertical cross-sectional geometry whose dimensions are 762 meters long by 7.62 meters wide by 15.24 meters thick, with $100\times 1\times 20$ computational cells. The permeability field of the model is shown in Fig. \ref{fig:perm_SPE10}, where the x-axis is plotted using a scale of 0.1 in order to obtain an easily readable representation. 

An injection-production scenario is considered, where the gas is injected at the left side of the domain (fully saturated with oil), and oil is produced at the right side. Both phases are considered incompressible, while the formation has a compressibility of $c_v=6\times 10^{-10}\,\mbox{Pa}^{-1}$ modeled by the linear method with a reference pressure of 689.476\,kPa, which is the initial pressure at the top of the model at point 0.0\,m. The initial void fraction is $\alpha_v=0.2$.

\begin{figure}[htb!]
  \centering
  \includegraphics[width=8cm]{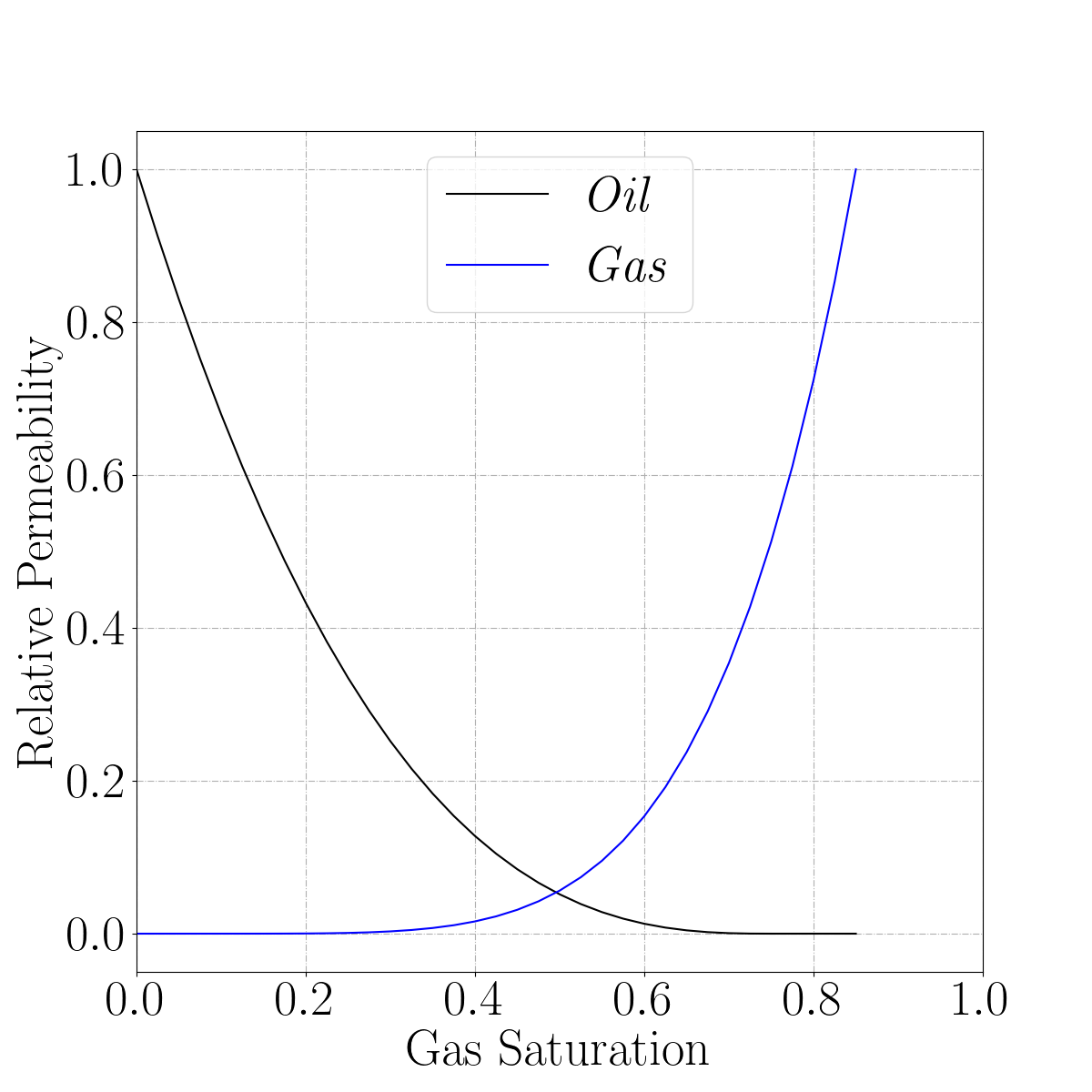}
  \caption{Relative permeabilities for the SPE10 project.}
  \label{fig:Rel_perm_SPE10}
\end{figure}

The model considers a residual oil fraction of 0.04, that is formulated as a stationary phase fraction in the OpenFOAM framework. Relative permeabilities, illustrated in Fig. \ref{fig:Rel_perm_SPE10}, are given by the authors in a table. The fluid properties are taken from the SPE10 dataset, that considers the densities $\rho_o=700\,\mbox{kg/m}^{3}$ and $\rho_g=1\,\mbox{kg/m}^{3}$, and viscosities $\mu_o=0.001\, \mbox{Pa}\cdot\mbox{s}$ and $\mu_g=0.00001\, \mbox{Pa}\cdot\mbox{s}$. 
A constant injection rate of $8.0671\times 10^{-5}\mbox{m}^3/\mbox{s}$ was applied, along with a production at a constant pressure of 655.002\,kPa. 
In this experiment, capillary effects are neglected, and the Courant time step control is selected with $C_{\max}=0.5$. 

\begin{figure}[htb!]
  \centering
  \includegraphics[width=13cm]{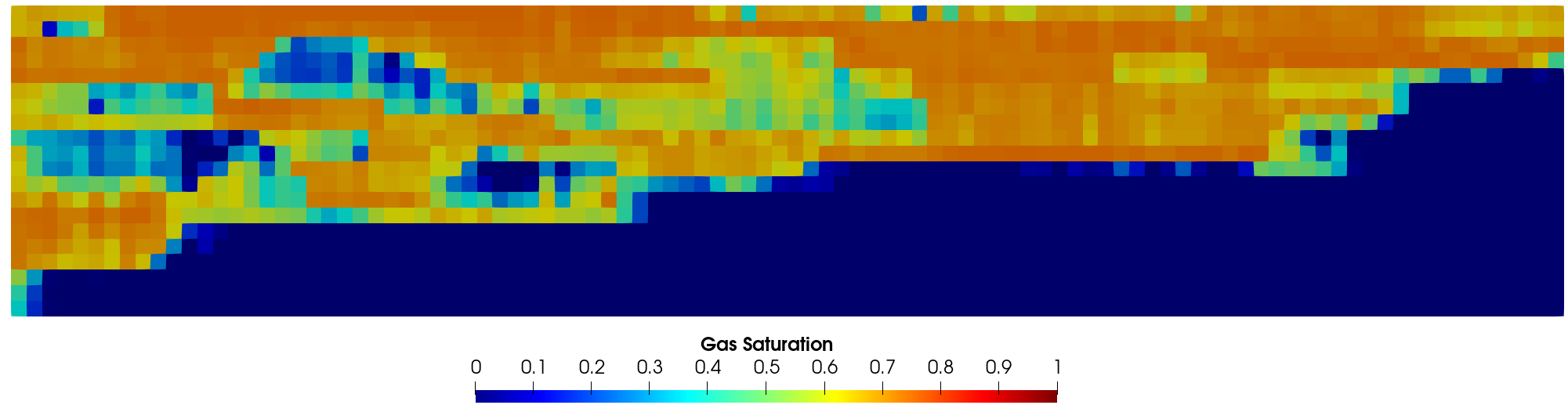}
  \caption{Gas saturation profile after 4800 days.}
  \label{fig:gas_sat_SPE10}
\end{figure}

The gas saturation profile after 4800 days can be seen in Fig. \ref{fig:gas_sat_SPE10}, where we can note how it flows through the heterogeneous medium. It is possible to observe that gas flows mainly in the superior part of the porous media. This behavior can be justified by the permeability field, presented in Fig. \ref{fig:perm_SPE10}.

\begin{figure}[htb!]
  \centering
  \includegraphics[width=7cm]{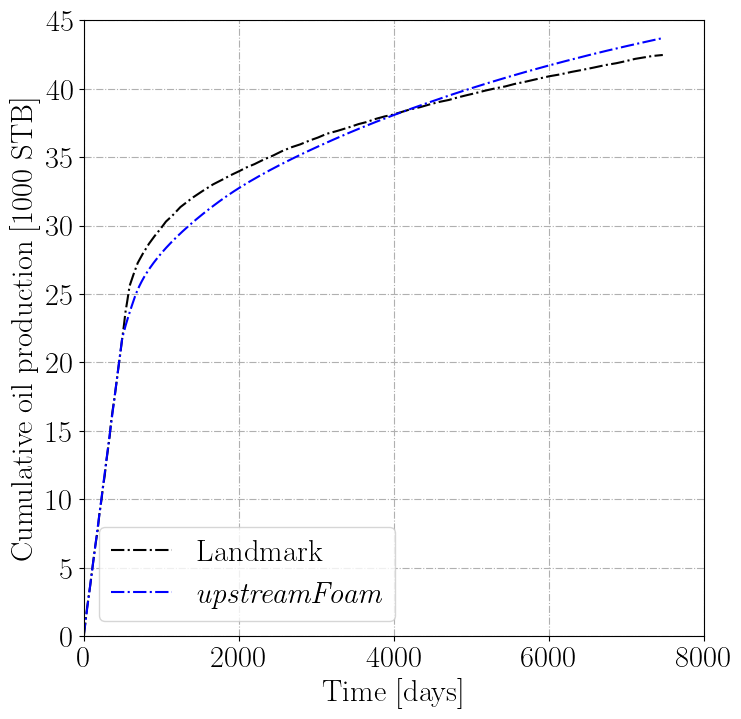}
  \caption{Cumulative oil production.}
  \label{fig:cumulative_oil_SPE10}
\end{figure}

Figure \ref{fig:cumulative_oil_SPE10} illustrates the cumulative oil production over a 21-year period (7665 days). The reference result from Landmark \citep{christie2001tenth}, obtained using the VIP simulator, is used for comparison. At the initial stages, \textit{upstreamFoam} exhibits excellent agreement with the reference result. However, after approximately 500 days, \textit{upstreamFoam} begins to underestimate the cumulative oil production. By 4000 days, it transitions to overestimating the production compared to Landmark's result. This discrepancy can be attributed to the well injector model, which applies a uniform flow rate across all corresponding cells. It is important to note that wellbore modeling falls outside the scope of this paper; however, it represents a potential avenue for further development and exploration by the authors. 
More specifically, ongoing work includes the implementation of a Peaceman well model that considers permeability in each cell and can induce actual reservoir productivity. This implementation will then serve as the basis for the development of improved models designed to simulate flows within wells with smart completions, taking into account localized head losses due to different equipment. With the new features, the solver will be able to generate enhanced predictions of well injectivity or productivity, impacting directly the practical decisions in production scenarios.

\subsection{Application in a 3D heterogeneous core}
\label{sub:multiphase_case_oil_gas_water_}

Numerical simulations of core flooding experiments are commonly used to validate measurements of rock properties. Therefore, the numerical section is closed with an application of the \textit{upstreamFoam} in a 3D heterogeneous core simulation.

\begin{figure}[htb!]
  \centering
  \includegraphics[width=12cm]{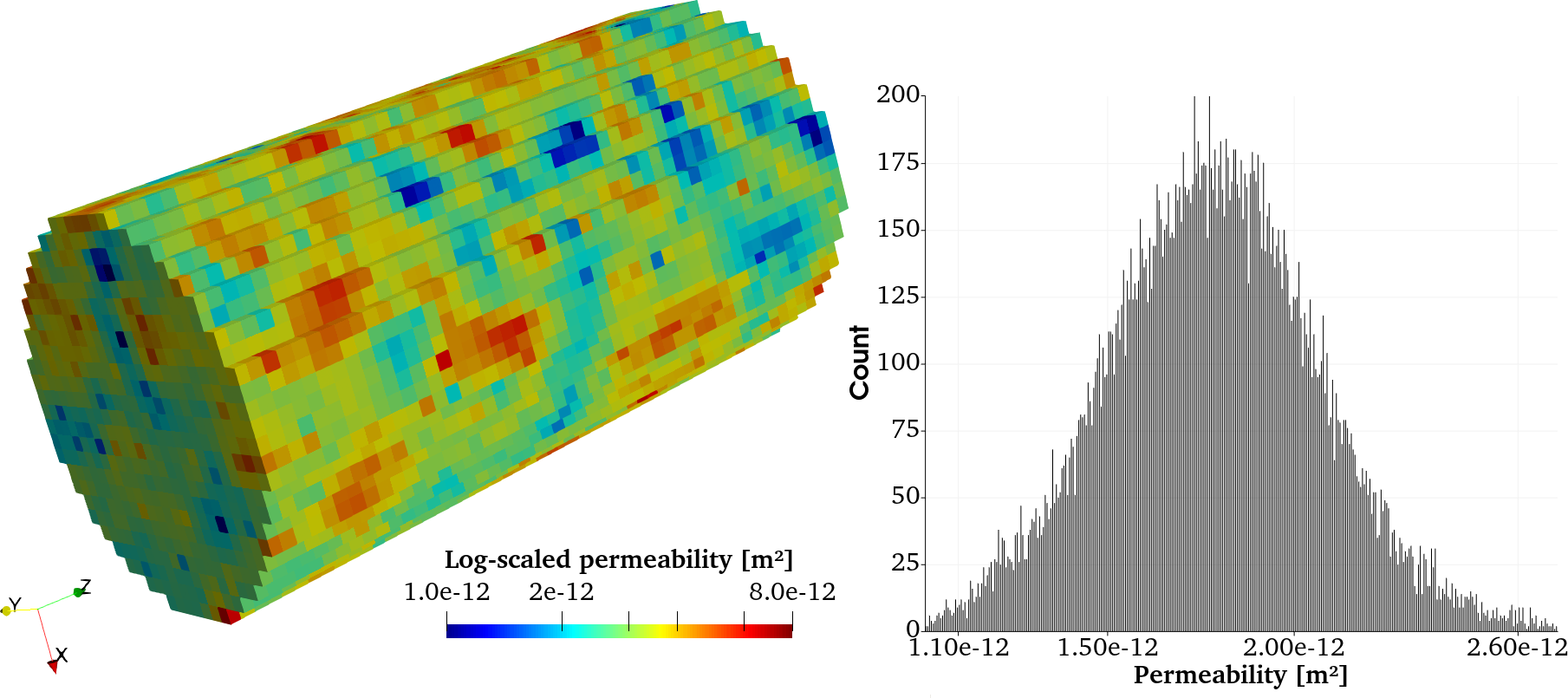}
  \caption{Permeability field of the manufactured core.}
  \label{fig:plug_perm}
\end{figure}

We consider a manufactured core with a relatively Gaussian distribution of the permeability field, as shown in Fig. \ref{fig:plug_perm}. The domain, with 25\,917 computational cells and dimensions of $0.05\,\mbox{m}\times0.05\,\mbox{m}\times0.106\,\mbox{m}$, has a homogeneous porosity of $\alpha_v=0.2$. The core is saturated with oil and water is injected from left at a constant volumetric rate of $1.66\times 10^{-8}\,\mbox{m/s}$, and the pressure at right is fixed in 0.1\, MPa.

In this experiment, gravity and compressibility are neglected. The phase densities are $\rho_w=1009\,\mbox{kg/m}^{3}$ and $\rho_o=800\,\mbox{kg/m}^{3}$, while the viscosities are $\mu_w=0.0005\, \mbox{Pa}\cdot\mbox{s}$ and $\mu_o=0.002\, \mbox{Pa}\cdot\mbox{s}$. We first consider a case without capillary pressure and set the Brooks and Corey relative permeability model with $n=1$ and $k_{r,i(\max)}=1$ for both phases. The Coats time step control with $C_{\max}=0.75$ is selected.

\begin{figure}[htb!]
  \centering
  \includegraphics[width=8cm]{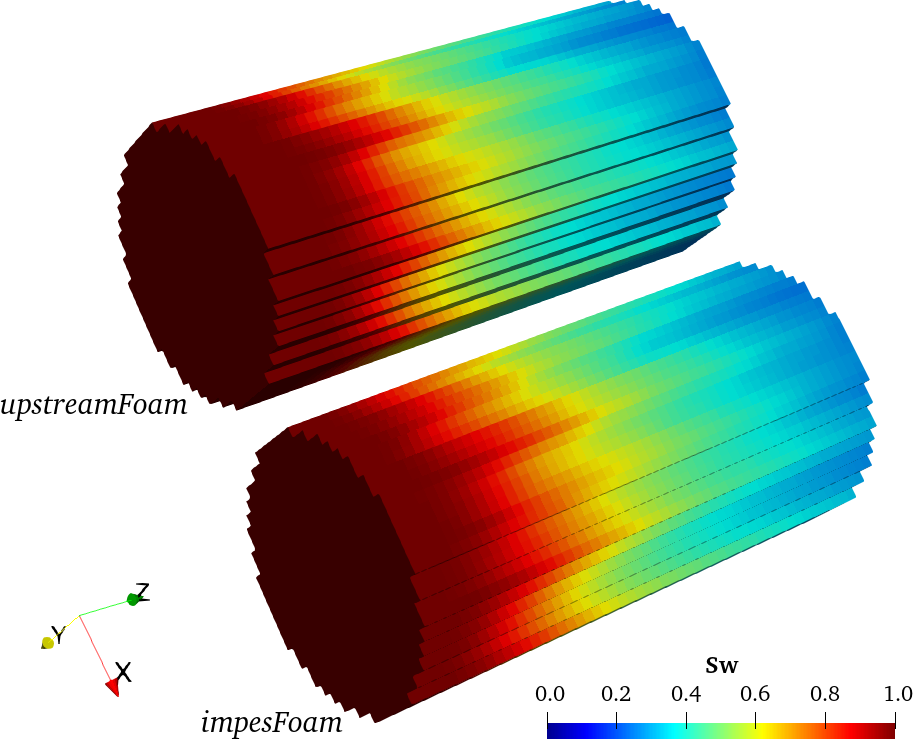}
  \caption{Water saturation profile after 2000\,s.}
  \label{fig:plug_sat_BC_1}
\end{figure}

The approximations provided by the \textit{upstreamFoam} and the reference solver \textit{impesFoam} \citep{horgue2015open} with an equivalent numerical setup are compared. Figure \ref{fig:plug_sat_BC_1} illustrates saturation profiles obtained by both solvers after 2000\,s, showing very similar behaviors. For this experiment, the computational time required by \textit{upstreamFoam} was 78.12\,s, while \textit{impesFoam} needed 30.11\,s.

\begin{figure}[htb!]
  \centering
  \includegraphics[width=11.5cm]{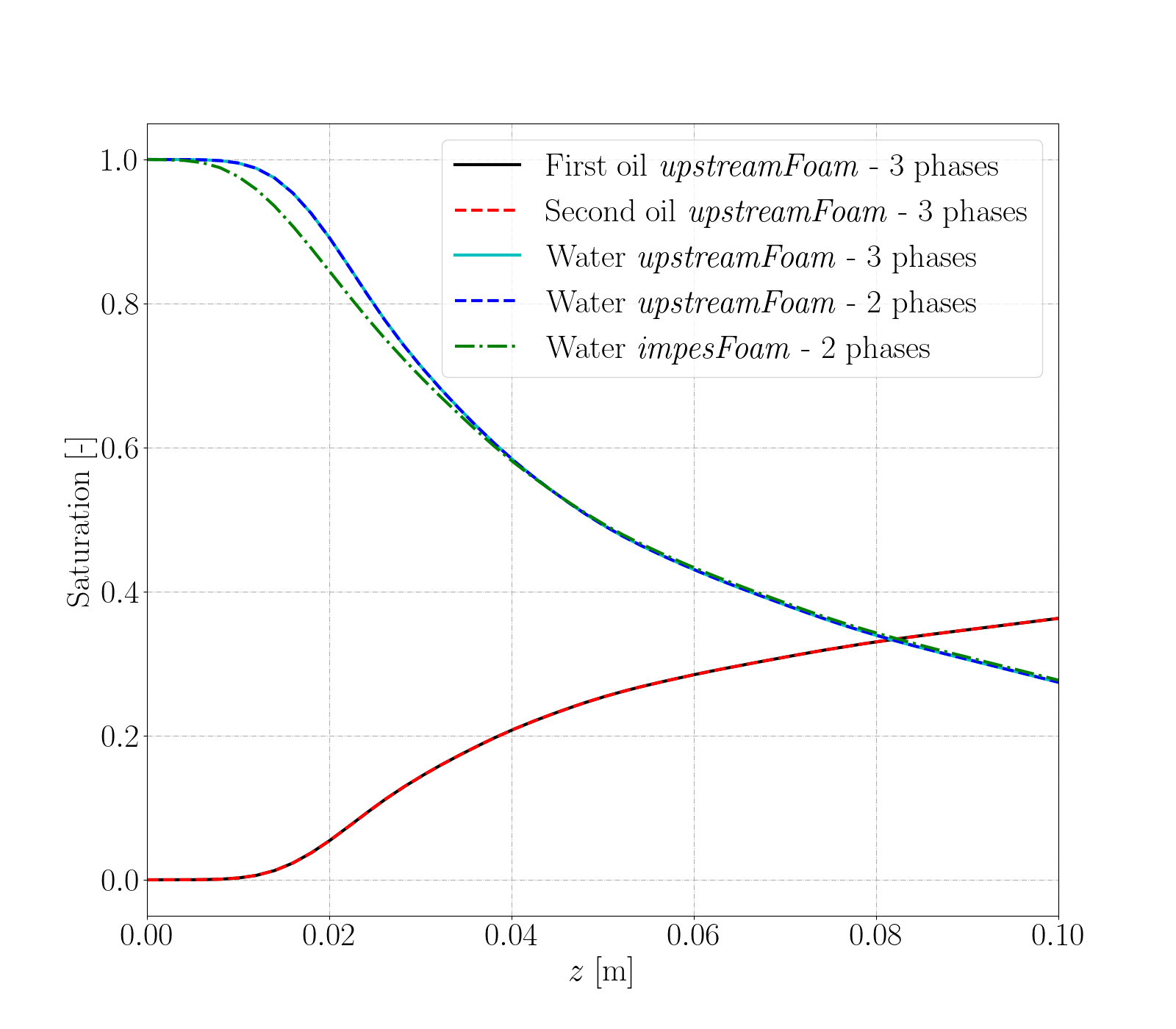}
  \caption{Saturations at 1000\,s in a center line across the core.}
  \label{fig:plug_line_BC_1}
\end{figure}

As an illustration of the applicability of the \textit{upstreamFoam} to more complex systems, we consider the previously test in a three-phase scenario with the oil divided into two phases. The core is initialized with saturation of 0.5 for each oil phase, that share the same fluid properties. The saturations at the center line across the core of the established three-phase flow and previous two-phase flow are presented in Fig. \ref{fig:plug_line_BC_1}. The \textit{upstreamFoam} produces the same water saturation for both two and three-phase flow regimes, demonstrating consistent results for the same flow condition, independent from the number of phases. We can also confirm that the saturations of both oil phases are equal. In the referred figure, the \textit{impesFoam} approximation for the two-phase flow is included, which is comparable to the \textit{upstreamFoam} solution. We emphasize that \textit{impesFoam} only considers two-phase flows, therefore, the same test with more mobile phases cannot be carried out on it. This study shows that our two-phase flow solution is in agreement with the produced by a reference solver as well as the capability of the \textit{upstreamFoam} to handle more mobile phases.

Our last experiment presents a qualitative study of the \textit{upstreamFoam} approximation for a case with capillary pressure and more realistic relative permeability curves by comparing our numerical results to the \textit{impesFoam} solution at equivalent conditions. We consider the same two-phase flow in the 3D core geometry and set the Brooks and Corey model for relative permeability with $n=2$ and $k_{r,i(\max)}=1$ for both phases. The Brooks and Corey model for the capillary pressure with $\beta=0.5$ and $p_{c,0}=10$\,Pa is chosen. 

\begin{figure}[htb!]
  \centering
  \subfigure[Saturation profiles  \label{fig:plug_sat_BC_2}]
  {\includegraphics[scale=0.2]{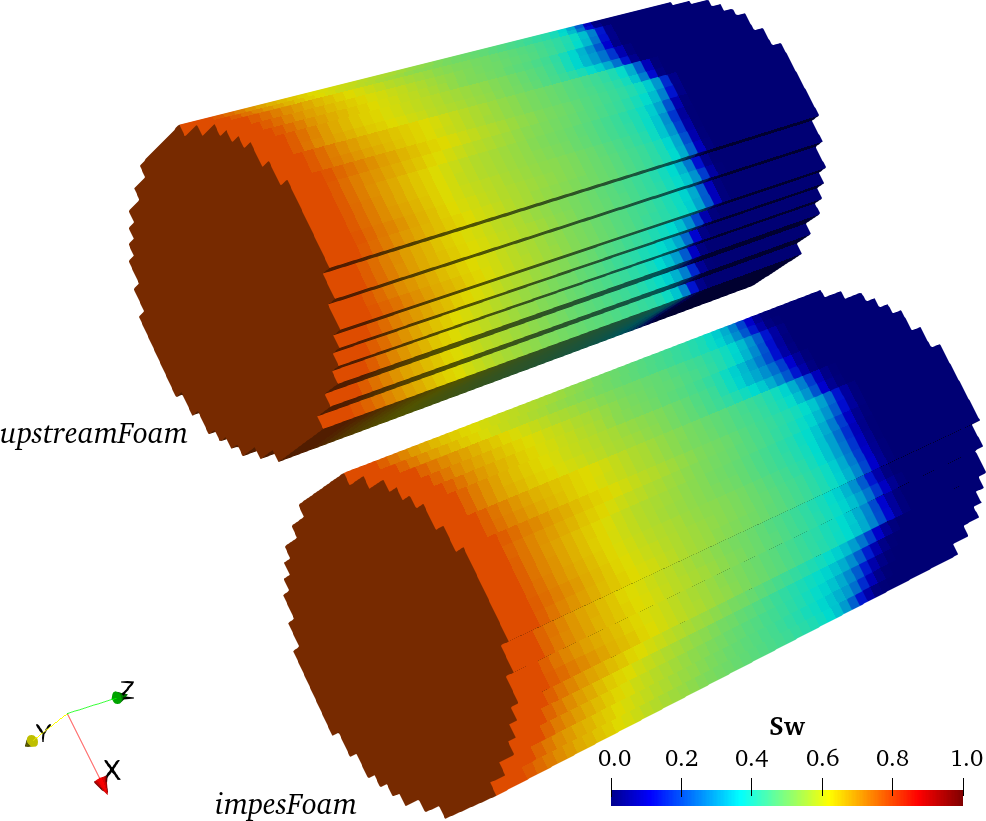}}\hspace{0.2cm}
  \subfigure[Time step history  \label{fig:plug_time_step}]
  {\includegraphics[scale=0.32]{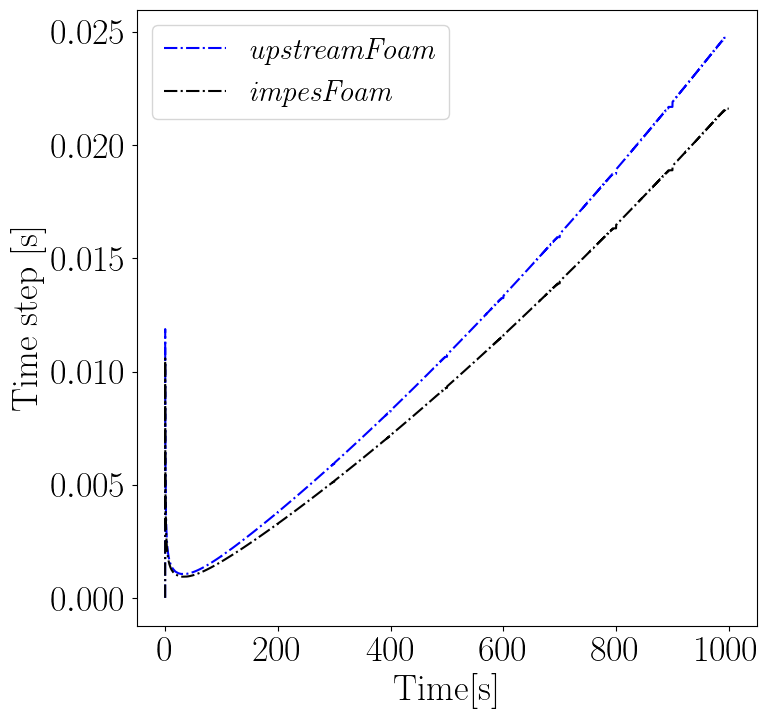}}
  \caption{Water saturation profile after 1000\,s and time step history for the case with capillary pressure and quadratic relative permeability curves.}
\end{figure}

Figure \ref{fig:plug_sat_BC_2} shows that the simulations performed with \textit{upstreamFoam} and \textit{impesFoam} develop similar saturation profiles. The computational times required by \textit{upstreamFoam} and \textit{impesFoam} were 5.99\,h and 2.29\,h, respectively. Again, it is important to highlight that despite the \textit{impesFoam} having a significantly lower computational cost, it presents a very limited number of applications when compared to \textit{upstreamFoam}.  The same time step Coats restriction has been set for both solvers, which presented similar time step histories, as reported in Fig. \ref{fig:plug_time_step}. We remark that the restricted time step sizes presented are close related to the capillary effects, that in this case are are not too strong due to the choice of $p_{c,0}$. Other temporal discretizations, such as implicit treatments, can be considered to overcome such limitations.

\section{Conclusions}%
\label{sec:conclusion}

In this work, we introduced and tested a new OpenFOAM application to simulate multiphase flows in porous media. The solver, called \textit{upstreamFoam}, combines the Eulerian multi-fluid formulation for a system of phase fractions with Darcy’s law for flows through porous media. It is based on the \textit{multiphaseEulerFoam} and includes models for reservoir simulation of the \textit{porousMultiphaseFoam}, taking advantage of the most recent technologies developed for these well-established solvers. 

The \textit{upstreamFoam} has been successfully tested in a wide range of multiphase flows in porous media. We verified the solver simulating classical problems with analytical, semi-analytical, and reference solutions given by the \textit{porousMultiphaseFoam}. Studies with different data and geometries have been performed to evaluate reservoir properties estimates, resulting in satisfactory accuracy. We also have demonstrated the potential of the solver to approximate complex problems in a multiphase scenario with more than two mobile phases.

The presented solver was developed in a general formulation, that can be extended to multiple species, mass transfer, and others complexities employing the tool-set, capabilities, and efficiency offered by the OpenFOAM framework. In future works, we intend to prepare the code for public release, include wellbore models, and investigate the use of new schemes to overcome the time step restrictions.

\section*{Acknowledgments}

The authors would like to thank PETROBRAS (grant 2017/00610-4) for the financial and material support.

%

 \bibliographystyle{elsarticle-harv} 
 \bibliography{references}

\begin{thebibliography}{48}
\expandafter\ifx\csname natexlab\endcsname\relax\def\natexlab#1{#1}\fi
\expandafter\ifx\csname url\endcsname\relax
  \def\url#1{\texttt{#1}}\fi
\expandafter\ifx\csname urlprefix\endcsname\relax\def\urlprefix{URL }\fi

\bibitem[{Aziz(1979)}]{aziz1979petroleum}
Aziz, K., 1979. Petroleum reservoir simulation. Applied Science Publishers 476.

\bibitem[{Bear(1988)}]{bear1988dynamics}
Bear, J., 1988. Dynamics of fluids in porous media. Courier Corporation, New
  York.

\bibitem[{Brinkman(1949)}]{brinkman1949calculation}
Brinkman, H.~C., 1949. A calculation of the viscous force exerted by a flowing
  fluid on a dense swarm of particles. Flow, Turbulence and Combustion 1~(1),
  27--34.

\bibitem[{Brooks(1965)}]{brooks1965hydraulic}
Brooks, R.~H., 1965. Hydraulic properties of porous media. Ph.D. thesis.

\bibitem[{Carrillo et~al.(2020)Carrillo, Bourg, and
  Soulaine}]{carrillo2020multiphase}
Carrillo, F.~J., Bourg, I.~C., Soulaine, C., 2020. Multiphase flow modeling in
  multiscale porous media: {A}n open-source micro-continuum approach. Journal
  of Computational Physics: X 8, 100073.

\bibitem[{Chen et~al.(2006)Chen, Huan, and Ma}]{chen2006computational}
Chen, Z., Huan, G., Ma, Y., 2006. Computational methods for multiphase flows in
  porous media. SIAM, Philadelphia, PA.

\bibitem[{Chen and Zhang(2009)}]{chen2009well}
Chen, Z., Zhang, Y., 2009. Well flow models for various numerical methods.
  International Journal of Numerical Analysis \& Modeling 6~(3).

\bibitem[{Christie and Blunt(2001)}]{christie2001tenth}
Christie, M.~A., Blunt, M.~J., 2001. Tenth {SPE} comparative solution project:
  A comparison of upscaling techniques. SPE Reservoir Evaluation \& Engineering
  4~(04), 308--317.

\bibitem[{Coats(2000)}]{coats2000note}
Coats, K.~H., 2000. A note on {IMPES} and some {IMPES}-based simulation models.
  SPE Journal 5~(03), 245--251.

\bibitem[{Coats(2003)}]{coats2003impes}
Coats, K.~H., 2003. {IMPES} stability: {S}election of stable timesteps. SPE
  Journal 8~(02), 181--187.

\bibitem[{Courant et~al.(1928)Courant, Friedrichs, and
  Lewy}]{courant1928partiellen}
Courant, R., Friedrichs, K., Lewy, H., 1928. {\"U}ber die partiellen
  {Differenzengleichungen} der mathematischen {Physik}. Mathematische annalen
  100~(1), 32--74.

\bibitem[{Crowe et~al.(2012)Crowe, Schwarzkopf, Sommerfeld, and
  Tsuji}]{crowe1998multiphase}
Crowe, C.~T., Schwarzkopf, J.~D., Sommerfeld, M., Tsuji, Y., 2012. Multiphase
  flows with droplets and particles. CRC Press - Taylor \& Francis Group, Boca
  Raton, FL.

\bibitem[{Dami{\'a}n and Nigro(2014)}]{damian2014extended}
Dami{\'a}n, S.~M., Nigro, N.~M., 2014. An extended mixture model for the
  simultaneous treatment of small-scale and large-scale interfaces.
  International Journal for Numerical Methods in Fluids 75~(8), 547--574.

\bibitem[{Drew and Passman(2006)}]{drew2006theory}
Drew, D.~A., Passman, S.~L., 2006. Theory of multicomponent fluids. Vol. 135.
  Springer Science \& Business Media, Berlin.

\bibitem[{Favero et~al.(2015)Favero, Silva, and Lage}]{favero2015modeling}
Favero, J.~L., Silva, L. F.~L., Lage, P.~L., 2015. Modeling and simulation of
  mixing in water-in-oil emulsion flow through a valve-like element using a
  population balance model. Computers \& Chemical Engineering 75, 155--170.

\bibitem[{Fioroni et~al.(2021)Fioroni, Larreteguy, and
  Savioli}]{fioroni2021openfoam}
Fioroni, S., Larreteguy, A.~E., Savioli, G.~B., 2021. An {OpenFOAM} application
  for solving the black oil problem. Mathematical Models and Computer
  Simulations 13~(5), 907--918.

\bibitem[{Forchheimer(1901)}]{ph1901wasserbewegung}
Forchheimer, P., 1901. Wasserbewegung durch boden. Zeitschrift des Vereines
  Deutscher Ingenieure 45~(50), 1781--1788.

\bibitem[{Franc et~al.(2016)Franc, Horgue, Guibert, and
  Debenest}]{franc2016benchmark}
Franc, J., Horgue, P., Guibert, R., Debenest, G., 2016. Benchmark of different
  {CFL} conditions for {IMPES}. Comptes Rendus M{\'e}canique 344~(10),
  715--724.

\bibitem[{Galeazzo et~al.(2024{\natexlab{a}})Galeazzo, Garcia-Gasulla, Boella,
  Pocurull, Lesnik, Rusche, Bn{\`a}, Cerminara, Brogi, Marchetti, Gregori,
  Wei$\beta$, and Ruopp}]{galeazzo2024performance}
Galeazzo, F. C.~C., Garcia-Gasulla, M., Boella, E., Pocurull, J., Lesnik, S.,
  Rusche, H., Bn{\`a}, S., Cerminara, M., Brogi, F., Marchetti, F., Gregori,
  D., Wei$\beta$, R.~G., Ruopp, A., 2024{\natexlab{a}}. Performance comparison
  of {CFD} microbenchmarks on diverse {HPC} architectures. Computers 13~(5),
  115.

\bibitem[{Galeazzo et~al.(2024{\natexlab{b}})Galeazzo, Wei$\beta$, Lesnik,
  Rusche, and Ruopp}]{galeazzo2024understanding}
Galeazzo, F. C.~C., Wei$\beta$, R.~G., Lesnik, S., Rusche, H., Ruopp, A.,
  2024{\natexlab{b}}. Understanding superlinear speedup in current {HPC}
  architectures. Preprints 2024040219.

\bibitem[{Goyeau et~al.(2003)Goyeau, Lhuillier, Gobin, and
  Velarde}]{goyeau2003momentum}
Goyeau, B., Lhuillier, D., Gobin, D., Velarde, M., 2003. Momentum transport at
  a fluid--porous interface. International Journal of Heat and Mass Transfer
  46~(21), 4071--4081.

\bibitem[{Greenshields and Weller(2022)}]{greenshields2022notes}
Greenshields, C.~J., Weller, H.~G., 2022. Notes on Computational Fluid
  Dynamics: General Principles. CFD Direct Ltd, Reading, UK.

\bibitem[{Hill(1998)}]{hill1998computer}
Hill, D.~P., 1998. The computer simulation of dispersed two-phase flow. Ph.D.
  thesis, University of London.

\bibitem[{Horgue et~al.(2022)Horgue, Renard, Gerlero, Guibert, and
  Debenest}]{horgue2022porousmultiphasefoam}
Horgue, P., Renard, F., Gerlero, G.~S., Guibert, R., Debenest, G., 2022.
  porousmultiphasefoam v2107: An open-source tool for modeling
  saturated/unsaturated water flows and solute transfers at watershed scale.
  Computer Physics Communications 273, 108278.

\bibitem[{Horgue et~al.(2015)Horgue, Soulaine, Franc, Guibert, and
  Debenest}]{horgue2015open}
Horgue, P., Soulaine, C., Franc, J., Guibert, R., Debenest, G., 2015. An
  open-source toolbox for multiphase flow in porous media. Computer Physics
  Communications 187, 217--226.

\bibitem[{Ishii and Hibiki(2010)}]{ishii2010thermo}
Ishii, M., Hibiki, T., 2010. Thermo-fluid dynamics of two-phase flow. Springer
  Science \& Business Media, New York.

\bibitem[{Jasak(1996)}]{jasak1996error}
Jasak, H., 1996. Error analysis and estimation for the finite volume method
  with applications to fluid flows. Ph.D. thesis, Imperial College London
  (University of London).

\bibitem[{Keser et~al.(2021)Keser, Battistoni, Im, and
  Jasak}]{keser2021eulerian}
Keser, R., Battistoni, M., Im, H.~G., Jasak, H., 2021. A {E}ulerian multi-fluid
  model for high-speed evaporating sprays. Processes 9~(6), 941.

\bibitem[{Leverett(1941)}]{leverett1941capillary}
Leverett, M., 1941. Capillary behavior in porous solids. Transactions of the
  AIME 142~(01), 152--169.

\bibitem[{Morel-Seytoux et~al.(1996)Morel-Seytoux, Meyer, Nachabe, Tourna,
  Van~Genuchten, and Lenhard}]{morel1996parameter}
Morel-Seytoux, H.~J., Meyer, P.~D., Nachabe, M., Tourna, J., Van~Genuchten,
  M.~T., Lenhard, R.~J., 1996. {Parameter equivalence for the Brooks-Corey and
  van Genuchten soil characteristics: Preserving the effective capillary
  drive}. Water Resources Research 32~(5), 1251--1258.

\bibitem[{Muskat(1981)}]{muskat1981physical}
Muskat, M., 1981. Physical principles of oil production. McGraw-Hill, New York.

\bibitem[{Nguyen et~al.(2022)Nguyen, Jung, Shim, and Yoo}]{nguyen2022real}
Nguyen, D.~N., Jung, K.~S., Shim, J.~W., Yoo, C.~S., 2022. Real-fluid
  thermophysicalmodels: An {OpenFOAM}-based library for reacting flow
  simulations at high pressure. Computer Physics Communications 273, 108264.

\bibitem[{Peaceman(1978)}]{peaceman1978interpretation}
Peaceman, D.~W., 1978. Interpretation of well-block pressures in numerical
  reservoir simulation (includes associated paper 6988). Society of Petroleum
  Engineers Journal 18~(03), 183--194.

\bibitem[{Rudman(1997)}]{rudman1997volume}
Rudman, M., 1997. Volume-tracking methods for interfacial flow calculations.
  International journal for numerical methods in fluids 24~(7), 671--691.

\bibitem[{Rusche(2003)}]{rusche2003computational}
Rusche, H., 2003. Computational fluid dynamics of dispersed two-phase flows at
  high phase fractions. Ph.D. thesis, Imperial College London (University of
  London).

\bibitem[{Sangnimnuan et~al.(2021)Sangnimnuan, Li, and
  Wu}]{sangnimnuan2021development}
Sangnimnuan, A., Li, J., Wu, K., 2021. Development of coupled two phase flow
  and geomechanics model to predict stress evolution in unconventional
  reservoirs with complex fracture geometry. Journal of Petroleum Science and
  Engineering 196, 108072.

\bibitem[{Silva et~al.(2008)Silva, Damian, and Lage}]{silva2008implementation}
Silva, L. F.~L., Damian, R., Lage, P.~L., 2008. Implementation and analysis of
  numerical solution of the population balance equation in {CFD} packages.
  Computers \& Chemical Engineering 32~(12), 2933--2945.

\bibitem[{Silva and Lage(2011)}]{silva2011development}
Silva, L. F.~L., Lage, P.~L., 2011. Development and implementation of a
  polydispersed multiphase flow model in {OpenFOAM}. Computers \& chemical
  engineering 35~(12), 2653--2666.

\bibitem[{Soulaine and Tchelepi(2016)}]{soulaine2016micro}
Soulaine, C., Tchelepi, H.~A., 2016. Micro-continuum approach for pore-scale
  simulation of subsurface processes. Transport in porous media 113, 431--456.

\bibitem[{Spalding(1981)}]{spalding1981numerical}
Spalding, D.~B., 1981. Numerical computation of multi-phase fluid flow and heat
  transfer. In Von Karman Inst. for Fluid Dyn. Numerical Computation of
  Multi-Phase Flows, 161--191.

\bibitem[{Sugumar et~al.(2020)Sugumar, Kumar, and
  Govindarajan}]{sugumar2020grid}
Sugumar, L., Kumar, A., Govindarajan, S.~K., 2020. Grid adaptation of
  multiphase fluid flow solver in porous medium by {OpenFOAM}. Petroleum \&
  Coal 62~(4).

\bibitem[{Tocci(2016)}]{tocci2016assessment}
Tocci, F., 2016. Assessment of a hybrid {VOF} two-fluid {CFD} solver for
  simulation of gas-liquid flows in vertical pipelines in {OpenFOAM}. Master's
  thesis, Politecnico di Milano, Italy.

\bibitem[{Wang et~al.(2018)Wang, Wang, Gao, Zhou, and
  Zhai}]{wang2018literature}
Wang, H., Wang, H., Gao, F., Zhou, P., Zhai, Z.~J., 2018. Literature review on
  pressure--velocity decoupling algorithms applied to built-environment {CFD}
  simulation. Building and Environment 143, 671--678.

\bibitem[{Wardle and Weller(2013)}]{wardle2013hybrid}
Wardle, K.~E., Weller, H.~G., 2013. Hybrid multiphase {CFD} solver for coupled
  dispersed/segregated flows in liquid-liquid extraction. International Journal
  of Chemical Engineering 2013.

\bibitem[{Weller(2002)}]{weller2002derivation}
Weller, H.~G., 2002. Derivation, modelling and solution of the conditionally
  averaged two-phase flow equations. Nabla Ltd, No Technical Report TR/HGW 2,
  9.

\bibitem[{Weller(2008)}]{weller2008new}
Weller, H.~G., 2008. A new approach to {VOF}-based interface capturing methods
  for incompressible and compressible flow. OpenCFD Ltd., Report TR/HGW 4, 35.

\bibitem[{Weller et~al.(1998)Weller, Tabor, Jasak, and
  Fureby}]{weller1998tensorial}
Weller, H.~G., Tabor, G., Jasak, H., Fureby, C., 1998. A tensorial approach to
  computational continuum mechanics using object-oriented techniques. Computers
  in physics 12~(6), 620--631.

\bibitem[{Wu(2015)}]{wu2015multiphase}
Wu, Y.-S., 2015. Multiphase fluid flow in porous and fractured reservoirs. Gulf
  professional publishing, Waltham, MA.

\end{thebibliography}





\end{document}